\documentclass[a4paper, 11pt]{article}
\usepackage{graphicx}
\usepackage{jheppub}
\usepackage[utf8]{inputenc}
\usepackage{yhmath}
\usepackage{caption}
\usepackage{subcaption}
\usepackage{hyperref}
\usepackage{enumerate}
\usepackage{wrapfig}
\usepackage{amsmath}
\usepackage{amsfonts}
\usepackage{booktabs}
\usepackage{comment}
\usepackage{cancel}
\usepackage{vmargin}
\usepackage[normalem]{ulem}
\setmargins{2.5cm}{1.5cm}{16.5cm}{23.42cm}{10pt}{1cm}{0pt}{2cm}
\usepackage{multirow}
\usepackage{slashed}
\usepackage{braket}
\usepackage{amssymb}
\usepackage{setspace}
\usepackage{amssymb}
\usepackage{url}
\usepackage{xcolor}
\usepackage{xspace}
\usepackage{cleveref}
\usepackage{nicematrix}
\DeclareMathOperator\arctanh{arctanh}
\usepackage[compat=1.1.0]{tikz-feynman}

   \title{\quad\quad\quad From \boldmath $B_c$ mesons to the baryon asymmetry:\\ \quad\quad\quad\quad\quad a unified $B$ Mesogenesis Framework} 
\author[a,b]{M. Burgos Marcos,}
\emailAdd{m.burgosmarcos@maastrichtuniversity.nl}
\author[a]{A. Verheyden,}
\emailAdd{a.verheyden@alumni.maastrichtuniversity.nl}
\author[a,b]{K. K. Vos}
\emailAdd{k.vos@maastrichtuniversity.nl}

\affiliation[a]{Gravitational Waves and Fundamental Physics (GWFP), Maastricht University, Duboisdomein 30, NL-6229 GT Maastricht, the Netherlands}
\affiliation[b]{Nikhef, Science Park 105, NL-1098 XG Amsterdam, the Netherlands}

\abstract{
    $B$ mesogenesis offers an interesting mechanism to generate the baryon asymmetry of the universe by converting the CP violation of the Standard Model into a net baryon number asymmetry. In this work we extend the $B$ mesogenesis framework by incorporating all relevant $B$ meson channels active after low-temperature reheating. We first revisit neutral $B$ mesogenesis, providing the time-dependent decay rates including electron scattering. We then perform a systematic study of $B_c$ decays, which are basically unexplored. We provide branching-ratio predictions using both leading-order factorization and a data-driven approach inspired by $D\to M_1M_2$ decays. These estimates allow us to quantify two $B_c$ sources of mesogenesis: the previously discussed $B_c^+ \to B^+ M^0$ channel and a new mechanism introduced in this work, $B_c^+ \to B_q^0 M$, in which the asymmetry is generated by combining direct CP violation with neutral-meson oscillations. Interestingly, in these channels the charm quark decays. Therefore, these decays give access to {\it charm CP violation} in modes with percent level branching ratios. With moderate assumptions, we find that $B_c$ mesogenesis can match or exceed the neutral contribution depending on the semileptonic asymmetry and reheating temperature. 
    Finally, we combine all three mechanisms and explore their viability in terms of the direct CP asymmetry, neutral-meson mixing parameters, early-universe fragmentation fractions and reheating temperature. We find that successful $B$ mesogenesis requires an enhancement in the fragmentation fraction and semileptonic asymmetry of $B_s^0$  at low reheating temperatures. For higher reheating temperatures, $B_c$ mesogenesis dominates and can explain the baryon asymmetry with fragmentation fraction and direct CP asymmetry of $B_c$ mesons at the percent level. Future measurements of $B_c$ modes are thus highly anticipated to further probe the viability of unified $B$ mesogenesis. 
}

\preprint{%
    Nikhef 2026-001
}

\begin{document}

\maketitle
\newpage
\spacing{1.35}

\section{Introduction}

One of the most striking features of our universe is the overwhelming dominance of matter over antimatter. High-precision measurements of the Cosmic Microwave Background and Big Bang Nucleosynthesis allow this imbalance to be quantified with remarkable accuracy \cite{Planck:2018vyg}:
\begin{equation}
    Y_\mathfrak{B}\equiv \frac{n_\mathfrak{B}-\bar n_\mathfrak{B}}{s}\approx 8.72 \times 10^{-11},
    \label{eq: baryon asymmetry}
\end{equation}
where $n_\mathfrak{B}$ and $n_\mathfrak{\bar B}$ are the baryon and antibaryon number densities, and $s$ is the entropy density of the universe. Despite its small numerical value, this asymmetry plays a decisive role in shaping the macroscopic structure of the universe, and its origin remains one of the central open questions at the interface of particle physics and cosmology. 

At first sight, such a tiny excess of matter over antimatter seems naively attainable within the Standard Model (SM), specially given the relatively large CP violation observed in certain processes, such as $B$ meson decays \cite{LHCb:2025cpvbar,Belle:2001zzw, Fleischer:2024uru}. However, no mechanism within the SM framework currently exists that can efficiently translate the CP violation present in the weak interaction into a net baryon asymmetry in the early universe. Moreover, according to the Sakharov conditions \cite{Sakharov:1967dj}, CP violation alone is insufficient to generate the observed baryon asymmetry: baryon number violation  and out of equilibrium conditions are also required. 

Several mechanisms have been proposed to explain the origin of the baryon asymmetry observed in the universe (see e.g.~\cite{Bodeker:2020ghk, Fukugita:1986hr, Barrow:2022gsu,vandeVis:2025efm}). In this work we focus on a promising candidate: $B$ mesogenesis \cite{Elor:2024cea,Elahi:2021jia, Nelson:2019fln, Elor:2018twp, Alonso-Alvarez:2021qfd}. This mechanism extends the SM particle content to simultaneously account for a visible baryon asymmetry and Dark Matter. The source of CP violation driving the baryon asymmetry arises from the SM itself. Previous studies have explored baryon asymmetry generation through $B_d^0-\bar B_d^0$ and $ B_s^0-\bar B_s^0$ oscillations \cite{Elor:2018twp}, and from CP violation in $B_c^+\to B^+$ decays \cite{Elahi:2021jia}. However, decays of $B_c^+ \to B_d^0 (B_s^0)$ can also contribute to the baryon asymmetry. In this work, we study this type of $B_c$ mesogenesis for the first time. We then combine all the potential contributions from $B$ decays to the baryon asymmetry. Note that, although the generation of the baryon asymmetry from $b$-baryon decays has also been explored in the literature~\cite{Aitken:2017wie}, we do not consider this mechanism in the present work.

Studying the viability of this unified $B$ mesogenesis framework depends on the SM properties of the $B$ mesons and their decays. Flavour transitions and CP violation have been extensively studied in the $B^+, B_d^0$ and $B_s^0$ sectors. In contrast, the heaviest meson in the family, $B_c$, remains relatively unexplored. 
Since $B_c$ decays are potentially crucial in the generation of baryon asymmetry within the mesogenesis framework, in this work we explore the two-body transitions $B_c^+ \to BM$ where $B=\{B^+,B_d^0,B_s^0\}$ and $M$ is a light pseudoscalar meson, $M=P\in\{\pi,K,\eta,\eta'\}$, or a light vector meson, $M=V\in\{\rho,K^*,\omega\}$, and evaluate their impact in baryon asymmetry. 

Experimental information on these channels is rather limited, mainly due to its much lower production rate at the Large Hadron Collider (LHC). To date, the only measured channel within this class is $B_c^+ \to B_s^0 \pi^+$, for which LHCb reports\footnote{Here we used the fragmentation fraction ratio measurement $\frac{f_c}{f_s}=0.029\pm 0.007$ \cite{LHCb:2019tea}.} \cite{LHCb:2013xlg, LHCb:2019tea,LHCb:2022htj}:
\begin{equation}
\mathcal{B}(B_c^+ \to B_s^0 \pi^+)=(8.2 \pm 2.3)\%\,.
\label{eq:Bcmeas}
\end{equation}
More measurements are expected to be available in the near future, partly driven by the increasing energies and luminosity at the LHC. 

The measured branching ratio of $B_c^+\to B_s^0\pi^+$ is among the largest measured in the $B$-meson sector, highlighting the phenomenological importance of $B_c$ decays and motivating a systematic analysis of the full set of $B_c^+ \to BM$ channels.

From the theoretical perspective, $B_c^+\to BM$ decays present several appealing features, even beyond the mesogenesis context. In these transitions, the $b$ quark acts as the spectator and the underlying flavour change is entirely driven by the $c$ quark. This feature is uncommon among heavy meson decays and turns the $B_c$ system into a complementary setting to study charm-mediated nonleptonic amplitudes and potential sources of charm CP violation.

The calculation of these decays is challenging due to their fully hadronic nature. As a consequence, there is currently no model-independent approach to evaluate the corresponding hadronic matrix elements. Previously, the branching ratios of these decays were estimated using factorization methods \cite{Sun:2008wa, Choi:2009ym} and studied under $SU(3)_F$ symmetry \cite{Bhattacharya:2017aao}.
In this work, we estimate the branching ratios of the $B_c^+\to B M$ decays within a factorization framework using both a leading-order factorization approach and data-driven estimates using measurements of $D\to M_1M_2$. 
These branching ratio predictions allow us to numerically assess, for the first time, the impact of $B_c$ mesons on the baryon asymmetry. We then discuss in detail the interplay between $B_c$ and neutral $B$ mesogenesis.

This paper is organized as follows: we start by reviewing the general framework of $B$ mesogenesis and present the relevant mechanisms contributing to the  baryon asymmetry. The charged $B_c^+ \to B^+$ channel is discussed in Sec.~\ref{sec:intro_chargedmeso}, the neutral $B_q^0$ mesons contributions in Sec.~\ref{sec:intro_neutmeso}, and the novel $B_c^+ \to B_q^0$ mechanism in Sec.~\ref{sec:intro_hybmeso}. In Sec.~\ref{sec:BcinSM} we study $B_c^+ \to BM$ decays within the SM using a factorization framework and obtain branching ratio estimates for these modes. In Sec.~\ref{sec:numerical_results}, we perform a numerical analysis of the individual contributions to the baryon asymmetry. Finally, we assess the viability of our unified $B$ mesogenesis in Sec.~\ref{sec: combined model}, exploring also the dependence on the input parameters. We conclude in Sec.~\ref{sec: conclusions}.

\section{\textbf{\textit{B}} mesogenesis: the general framework} \label{sec: B mesogenesis}
The starting point of $B$ mesogenesis is the introduction of some heavy scalar field, $\Phi$, with mass $M_\Phi \sim 10-100 \,\mathrm{GeV}$ \cite{Elahi:2021jia, Elor:2018twp}, which dominates the energy density of the universe after inflation. 
This scalar field will decay eventually into $b$ and $\bar b$ pairs. The temperature at which $\Phi$ decays is required to be cool enough $\sim\mathcal{O}(10-100 \,\mathrm{MeV})$ such that the quarks hadronize before they decay. While decays into heavier SM particles, such as the top quark or the Higgs boson, are kinematically allowed, these heavier states eventually decay into bottom quarks. Consequently, we can assume that the $\Phi$ particle decays $100\%$ to $b$ quarks. 

The $b$ quarks then hadronize into the full spectrum of neutral and charged pseudoscalar $B$ mesons\footnote{$B$-meson resonances may also be produced, although they are expected to rapidly decay into the pseudoscalar meson states. Therefore, it can be safely assumed that effectively only the pseudoscalar $B$ mesons are present.}, whose decays are the source of the visible baryon asymmetry in the mesogenesis framework. 

Before addressing how the baryon asymmetry itself is generated, we first summarize how baryons are produced in this framework. Although the SM allows for baryon–antibaryon production, any process that yields a net baryon number from meson decays necessarily requires an extension of the SM particle content. In the $B$ mesogenesis setup, the $B$ mesons decay into a SM baryon $\mathfrak{B}= p, \Sigma, \Xi,...$ inclusively, i.e. alongside any number of additional light mesons or other SM particles, and an unstable Dirac fermion $\psi$, mediated by a new scalar particle $Y$ that carries electric and colour charge. We denote this inclusive branching ratio with the shorthand notation $B\to \mathfrak{B} \psi$. These decays are CP conserving and flavour specific. The Dirac fermion $\psi$ subsequently decays into two dark-sector particles: a Majorana fermion $\xi$ and a stable scalar baryon $\phi$, which together could account for the dark matter abundance. This secondary decay is needed to prevent the visible baryon asymmetry from being washed out by inverse processes. The out-of-equilibrium nature of this decay chain and the violation of visible baryon number fulfill two of the necessary Sakharov conditions for successful baryogenesis.

\begin{figure}[t]
    \centering
    \includegraphics[width=0.35\textwidth]{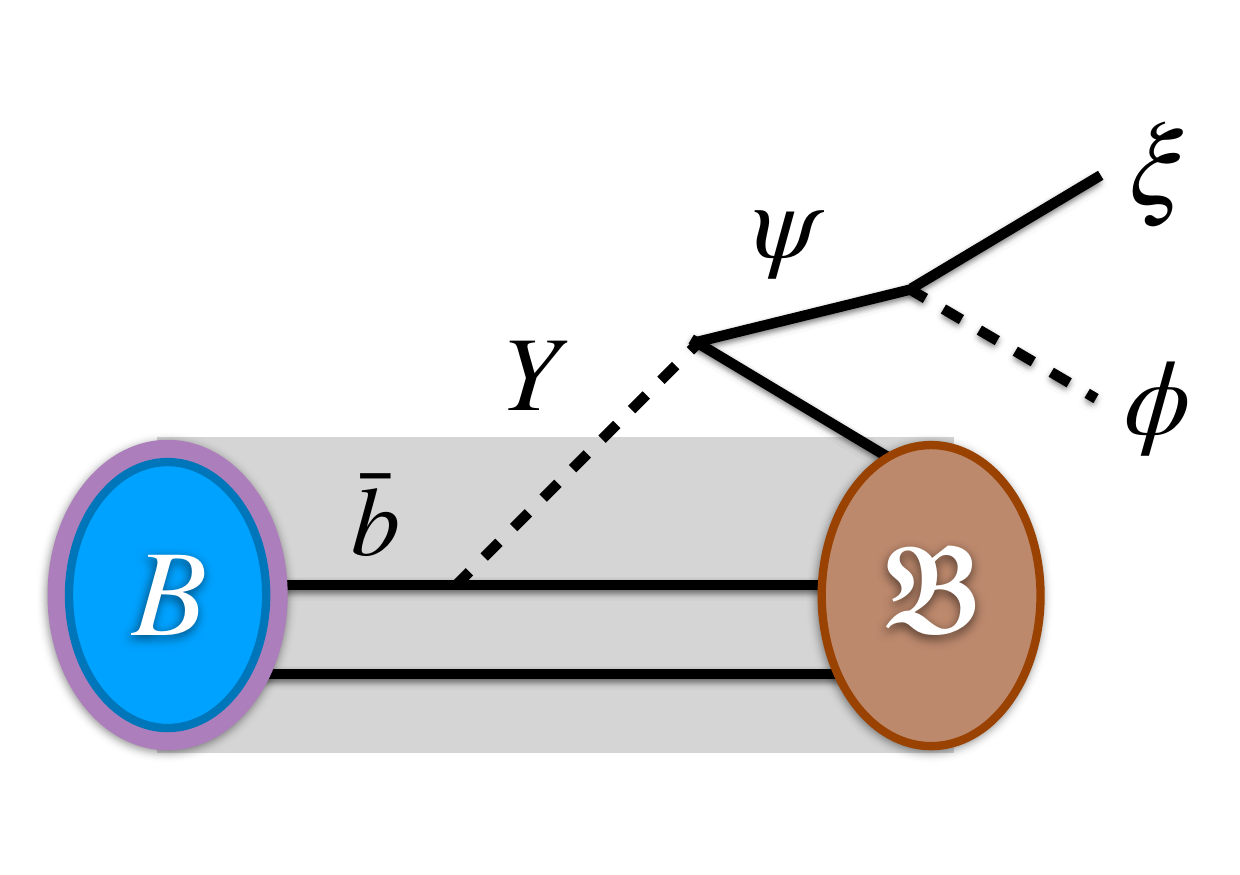}
    \caption{Illustration of the decay of a $B$ meson into a visible SM baryon $\mathfrak{B}$ and dark matter states $\xi$ and $\phi$, mediated by the heavy field $Y$.} 
    \label{fig:decay_diagram}
\end{figure}

Figure~\ref{fig:decay_diagram} shows a schematic diagram of the process that generates the SM baryons. For discussions on the dark-sector dynamics and the collider constraints on the new states and decay channels involved, we refer to, e.g., \cite{Berger:2024, Alonso-Alvarez:2021qfd, Shi:2023,Lenz:2024rwi}.

In this paper, we present a unified $B$ mesogenesis model, combining the effect of neutral $B_{s,d}^0$ and charged $B_c$ modes after their generation through the $\Phi$ field. Specifically, we define the total baryon asymmetry as:
\begin{equation}
    Y_\mathfrak{B}= f_q \;Y_\mathfrak{B}^{B^0_q}+f_c\left(\;Y_\mathfrak{B}^{B_c^+ \to B^+}+\; Y_\mathfrak{B}^{B_c^+ \to B_q^0}\right),
    \label{eq: individual Y's}
\end{equation}
where $f_q (f_c)$ is the fragmentation fraction defined as the probability that a $b$ quark hadronizes into a $B_q^0 (B_c^+)$ meson, and parametrizes the primordial abundance of the respective mesons. Each $Y_{\mathfrak{B}}^X$ is obtained by solving the Boltzmann equation \cite{Elahi:2021jia, Elor:2018twp}:
\begin{equation}
    \frac{d}{dt}(sY_\mathfrak{B}^X) + 3H(sY_\mathfrak{B}^X) =\frac{1}{2}\Gamma_\Phi^b n_\Phi \Delta_X.
    \label{eq: evolution individual baryon asymmetries}
\end{equation}
where $s$ denotes the entropy density of the universe and its determination and evolution can be found in \cite{Laine:2015kra}. The product $s Y_\mathfrak{B}^X = n_\mathfrak{B}^X-n_\mathfrak{\bar B}^X$, according to Eq.~\eqref{eq: baryon asymmetry}, represents the number density asymmetry of baryons generated through the specific mechanism $X \in \{B_q^0, B_c \to B^+, B_c^+ \to B_q^0\}$.
 A more detailed explanation on the numerical evaluation of $Y_\mathfrak{B}^X$ can be found in Appendix~\ref{ap: Boltzman eqs}. The parameter $\Delta_X$ quantifies the net baryon asymmetry generated by the different mesogenesis mechanisms. Since the $B\to \mathfrak{B}\phi \xi$ decay is CP-conserving it does not generate an asymmetry on its own. Any net baryon excess contributing to $\Delta_X$ must originate earlier, through an imbalance in the number of $B$ and $\bar B$ mesons prior to their decay. This imbalance arises from CP violation within the $B$ meson sector, through SM processes.

\begin{figure}[t]
    \centering
    \includegraphics[width=0.85\textwidth]{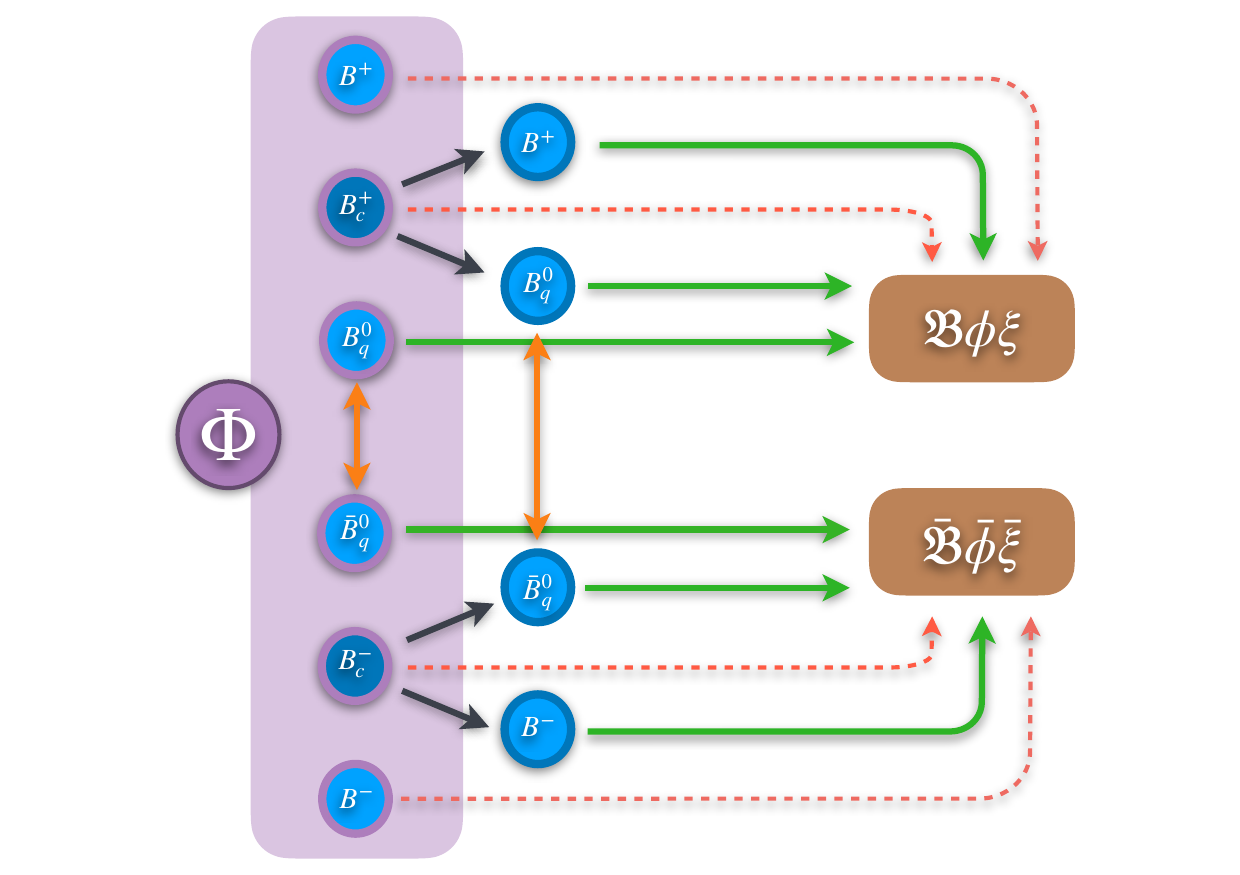}
    \caption{Schematic representation of the combined $B$ mesogenesis model. $B$ mesons in the purple-shaded area arise from hadronization of the $b$ quarks produced in $\Phi$ decays. Direct CP violation in decays is denoted by black arrows, while CP violation arising through $B_q^0$--$\bar{B}_q^0$ mixing is shown in orange. Green arrows indicate the decay paths responsible for generating the baryon asymmetry, whereas red dashed arrows denote channels that, although allowed, do not contribute to it.} 
    \label{fig: combined model}
\end{figure}

A schematic representation of the unified framework and the CP-violating processes contributing to the total asymmetry is shown in Fig.~\ref{fig: combined model}. Here the green arrows indicate decays that contribute to the baryon asymmetry.
 The $B$ meson asymmetry that generates $Y_\mathfrak{B}^{B_q^0}, \;\Delta_{B_q^0}$, originates from CP violation in neutral-meson mixing,  $B_q^0-\bar B_q^0$, with $q=d,s$, indicated by the orange arrows in Fig.~\ref{fig: combined model}. The generation of $\Delta_{B_q^0}$ is known as {\it neutral $B$ mesogenesis}. 
The asymmetry term $\Delta_{B_c^+\to B^+}$ is generated by direct CP violation in the decay of $B_c^+\to B^+$, marked by black arrows in Fig~\ref{fig: combined model}, and constitutes {\it charged $B_c$ mesogenesis}. These two processes were previously studied separately. In the following we revise their magnitude and study their impact on the total baryon asymmetry.
Finally, $\Delta_{B_c^+\to B_q^0}$ is a new contribution studied for the first time in this work: {\it hybrid $B_c$ mesogenesis}. 
This mechanism proceeds in two steps: 
\begin{itemize}
    \item The first step is the $B_c^+ \to B^0_q$ decay. Depending on the decay structure, this process can be either CP conserving or CP violating. In the latter case, an imbalance between the daughter $B_q^0$ and $\bar B_q^0$ mesons is generated.

    \item Subsequently, the $B_q^0$ and $\bar B_q^0$ mesons oscillate, producing mixing-induced CP violation.
\end{itemize}
If the $B_c^+ \to B_q^0$ decay is CP conserving, the mechanism is analogous to neutral $B$ mesogenesis. However, if the initial decay is CP violating, both direct and mixing induced CP violation contribute  to $\Delta_{B_c^+ \to B_q^0}.$ In the following, we discuss the different $\Delta_X$ terms.

\subsection{Charged $\boldsymbol{B_c}$ Mesogenesis}\label{sec:intro_chargedmeso}
We first briefly discuss the contribution $\Delta_{B_c^+\to B^+}$. Here the net baryon asymmetry is created through direct CP violation in the weak $B_c^+\to B^+$ decay. Given the limited available phase space in this class of decays, together with the challenges of reliably estimating three-body decay amplitudes, we restrict our analysis to two-body decay modes with a light pseudoscalar or vector meson in the final state. 

The direct CP asymmetry is defined as 
\begin{equation}\label{eq:Adir}
    \mathcal{A}_\text{CP}^\text{dir}(B^+_c\to B^+M^0)=\frac{\Gamma(B^+_c\to B^+M^0) - \Gamma(B^-_c\to  B^-\bar M^0)}{\Gamma(B^+_c\to B^+M^0) + \Gamma( B^-_c\to  B^-\bar M^0)},
\end{equation}
while mixing-induced CP asymmetry is absent. The resulting baryon asymmetry is given by \cite{Elahi:2021jia}:
\begin{equation}\label{eq:chargedBc}
    \Delta_{B_c^+\to B^+}=\sum_{M^0}\mathcal{A}_\text{CP}^\text{dir}(B_c^+\to B^+M^0)\mathcal{B}(B_c^+ \to B^+M^0)\mathcal{B}(B^+\to f) \ ,
\end{equation}
where $f = \mathfrak{B}\xi \phi$. We defined the CP-averaged branching ratio as
\begin{equation}\label{eq:Brdef}
    \mathcal{B}(B_c^+\to B^+M^0) \equiv \frac{\tau_{B_c^+}}{16\pi\, m_{B_c^+}} \, \Phi\left(m_{B^+}/m_{B_c^+},m_{M^0}/m_{B_c^+}\right)\frac{\Gamma(B_c^+\to B^+M^0) + \Gamma(B_c^-\to B^- \bar M^0)}{2} 
\end{equation}
where the phase space function $\Phi(x,y)$ is:
\begin{equation}
\Phi(x,y)=\sqrt{\left[1-(x+y)^2\right]\left[1-(x-y)^2\right]}   \ .  
\end{equation}

The branching ratio $\mathcal{B}(B^+\to \mathfrak{B}\xi\phi)$ describes the decay of a $B^+$ meson into a SM baryon and dark-sector states. Currently, experimental searches at Belle II and BaBar place stringent constraints on exclusive modes, such as $\mathcal{B}(B \to p + \text{invisible})\lesssim 10^{-5}$ \cite{Belle:2026kum}. However, as discussed, the relevant mesogenesis observable is the inclusive decay rate. Recent studies using light-cone sum rules show that once all hadronic final states are summed over, the inclusive branching ratio is enhanced by two to three orders of magnitude compared to the exclusive limits \cite{Mohamed:2025zgx, Boushmelev:2023huu, Khodjamirian:2022vta,Lenz:2024rwi}. On the other hand, such inclusive rates are indirectly constrained by LHC searches for jets plus missing transverse energy, which set strong bounds on the underlying effective operators involving $b$ quarks \cite{Alonso-Alvarez:2021qfd}, as well as other baryon number violating operators \cite{Hiller:2026osz}. Taking these complementary limits into account, as well as the sum over all possible SM baryon final states rather than only protons, we adopt $\mathcal{B}(B\to \mathfrak{B}\xi\phi)\simeq 10^{-3}$ for all flavours of $B$ mesons. Since the baryon asymmetry depends linearly on this branching ratio, any future updates or revisions to its value will scale the resulting asymmetry accordingly.

Charged $B_c$ mesogenesis was previously studied \cite{Elahi:2021jia} but a numerical analysis was not performed. Since no measurements are available for this decay class, we provide predictions for $B_c^+ \to B^+ M^0$ observables, which allows us to evaluate the impact of these decays numerically. 

We note that the $B_c^+\to B^+$ is not the only proposed mechanism of charged $B$ mesogenesis. Additional sources of asymmetry may arise from decays of $B^+$ \cite{Elahi:2021jia} or $D^+$ \cite{Elor:2020tkc} mesons into charged light mesons, which subsequently decay into a SM lepton and a dark-sector lepton. In order to translate the generated SM lepton asymmetry into a net baryon asymmetry, this mechanism requires introducing additional particles. Therefore, we do not include this mechanism in our joint analysis.

\subsection{\boldmath Revisiting Neutral $B$ Mesogenesis} \label{sec:intro_neutmeso}
We revisit the baryon asymmetry generated through the CP-violating effects in the oscillations of neutral $B$ mesons, through the parameter $\Delta_{B_q^0}$.

We begin by recalling the mechanism of neutral-meson mixing, where the mass eigenstates, $B^0_{q,H}$ and $B^0_{q,L}$, arise as quantum superpositions of the flavour ones, $B_q^0$ and $\bar B_q^0$,
\begin{equation}
\begin{aligned}
    |B_{q,L}^0 \rangle &= \mathfrak{p}_q|B_q^0\rangle + \mathfrak{q}_q |\bar B_q^0\rangle, \\
    |B_{q,H}^0 \rangle &= \mathfrak{p}_q|B_q^0\rangle - \mathfrak{q}_q |\bar B_q^0\rangle.
    \end{aligned}
\end{equation}

It is precisely because the physical states are not diagonal in flavour that a neutral $B$ meson can oscillate between $B_q^0$ and $\bar B_q^0$, with CP violation manifesting in the process if $|\mathfrak{q}_q/\mathfrak{p}_q|\neq 1$. 

The contribution to the baryon asymmetry can be derived from the time-dependent decay rates. However, unlike the standard evolution in vacuum, the neutral mesons in the early Universe are produced within a thermal bath. Continuous interactions with the plasma components, predominantly elastic scattering with electrons, act as ``environmental'' measurements that induce quantum decoherence. To properly account for this effect, the system can be described from the evolution of the density matrix \cite{Alonso-Alvarez:2021qfd}:
\begin{equation}
    \frac{dn}{dt}= -i[\mathcal H,n]-\frac{1}{2}\Gamma_{eB}[O_-,[O_-,n]]\quad \mathrm{with} \quad n=\left( \begin{aligned}
        n_{BB} \quad& n_{B\bar B} \\
        n_{\bar B B} \quad& n_{\bar B \bar B}
    \end{aligned}
            \right).
        \label{eq: density matrix}
\end{equation}
where $\mathcal H=M-\frac{i}{2} \Gamma$ is the effective Hamiltonian for $B$ meson oscillations, $M$ and $\Gamma$ the mass and decay rate matrices respectively, and $O_-=\text{diag}(1,-1)$. We use the density matrix formalism to track the evolution of the $B_q^0-\bar B_q^0$ system, where the diagonal entries $n_{BB},n_{\bar B \bar B}$ represent the number densities of $B_q^0$ and $\bar B_q^0$ mesons, respectively, and the off-diagonal terms represent the interference between the $B_q^0$ and $\bar B_q^0$ flavour states, which drive the meson oscillations.
The scattering rate of $B$ mesons to electrons is taken to be \cite{Elor:2018twp}:
\begin{equation}
    \Gamma_{eB}\simeq 10^{-11}\mathrm{GeV}\left( \frac{T}{20 \;\mathrm{MeV}}\right)^5 \left( \frac{\langle r_B^2 \rangle}{0.187\;\mathrm{fm}^2}\right)^2
    \label{eq: scattering electrons}
\end{equation}
where $T$ is the temperature of the plasma and $\langle r_B^2\rangle$ the charge radius of the $B$ meson \cite{Hwang:2001th}. In the following, we use $\langle r_B^2\rangle = 0.187$ fm$^2$.
 We define $A_f$ as the amplitude of the $B\to f$ decay and equivalently for $A_{\bar{f}}, \bar{A }_f$ and $\bar{A}_{\bar f}$, where in our case $f=\mathfrak{B}\xi\phi$. As the $B\to f$ decay is flavour specific, we have that $A_{\bar{f}}= \bar{A}_f =0$. 

Solving Eq.~\eqref{eq: density matrix} for flavour specific final states, the time-dependent decay rates are:
\begin{align}
    \Gamma(B_q^0(t) \to f) &= \mathcal{N}_f |A_f|^2 e^{-\tilde\Gamma_q t}\frac{1}{2}\left\{
 \frac{\Delta\Gamma_q}{\widetilde{\Delta\Gamma_q}}\cosh\left(\frac{\widetilde{\Delta \Gamma_q}\, t}{2} + \delta_{\Delta\Gamma_q}^\mathcal{S}\right)
+  \frac{\Delta M_q}{\widetilde{\Delta M_q}}\cos(\widetilde{\Delta M_q}\, t - \delta_{\Delta M_q}^\mathcal{S})
\right\} \label{eq: decay rate B to f},\\
 \Gamma(\bar B_q^0(t) \to \bar f) &= \mathcal{N}_f |\bar A_{\bar f}|^2 e^{-\tilde\Gamma_q t}
\frac{1}{2}\left\{ \frac{\Delta\Gamma_q}{\widetilde{\Delta\Gamma_q}}\cosh\left(\frac{\widetilde{\Delta \Gamma_q}\, t}{2} + \delta_{\Delta\Gamma_q}^\mathcal{S}\right)
+  \frac{\Delta M_q}{\widetilde{\Delta M_q}}\cos(\widetilde{\Delta M_q}\, t - \delta_{\Delta M_q}^\mathcal{S})
\right\} \label{eq: decay rate Bbar to fbar},\\
\Gamma(B_q^0(t) \to \bar f) &= \mathcal{N}_f |\bar A_{\bar f}|^2 e^{-\tilde\Gamma_q t}(1-a_q)\frac{1}{2}\left\{  \frac{\Delta\Gamma_q}{\widetilde{\Delta\Gamma_q}}\cosh\left(\frac{\widetilde{\Delta \Gamma_q}\, t}{2} + \delta_{\Delta\Gamma_q}^\mathcal{S}\right)
-  \frac{\Delta M_q}{\widetilde{\Delta M_q}}\cos(\widetilde{\Delta M_q}\, t - \delta_{\Delta M_q}^\mathcal{S})
\right\} \label{eq: decay rate B to fbar},\\
 \Gamma(\bar B_q^0(t) \to f) &= \mathcal{N}_f |A_f|^2 \frac{1}{1-a_q}e^{-\tilde\Gamma_q t}
\frac{1}{2}\left\{  \frac{\Delta\Gamma_q}{\widetilde{\Delta\Gamma_q}}\cosh\left(\frac{\widetilde{\Delta \Gamma_q}\, t}{2} + \delta_{\Delta\Gamma_q}^\mathcal{S}\right)
-  \frac{\Delta M_q}{\widetilde{\Delta M_q}}\cos(\widetilde{\Delta M_q}\, t - \delta_{\Delta M_q}^\mathcal{S}) \right\} \label{eq: decay rate Bbar to f},
\end{align}
where we have defined the ``effective'' mixing quantities: 
\begin{equation}
\begin{aligned}
    \widetilde{\Delta M_q}&=\sqrt{\Delta M_q^2 - \Gamma_{eB}^2}\,, \,&\widetilde{\Delta \Gamma_q}&=\sqrt{\Delta \Gamma_q^2 + 4 \Gamma_{eB}^2}\,, \, &\tilde \Gamma_q &= \Gamma_q + \Gamma_{eB}\,,\, \\
    \delta_{\Delta M_q}^\mathcal{S}&=\arctan\left(\frac{\Gamma_{eB}}{\widetilde{\Delta M_q}}\right)& \delta_{\Delta \Gamma_q}^\mathcal{S}&=\arctanh\left(\frac{2\Gamma_{eB}}{\widetilde{\Delta \Gamma_q}}\right)
    \end{aligned}
    \label{eq: effective mixing variables}
\end{equation}
which are given in terms of the standard mixing variables

$\Delta M_q=M_{q,H} - M_{q,L} \simeq 2 |M_{12}|$, $\Delta \Gamma_q = \Gamma_{q,L} - \Gamma_{q,H} \simeq 2|\Gamma_{12}|\cos\phi_{12} $. Here $\Gamma_q$ is the total decay rate of the neutral meson and $\mathcal{N}_f$ is a normalization factor. 
The effective mixing quantities in Eq.~\eqref{eq: effective mixing variables} depend on the plasma temperature through $\Gamma_{eB}$, as given in Eq.~\eqref{eq: scattering electrons}. The standard, vacuum mixing parameters are recovered in the $T,\Gamma_{eB}\to 0$ limit.

In addition, we define \cite{Nierste:2009wg}
\begin{equation}\label{eq:aqdef}
a_q \equiv 1 - \left|\frac{\mathfrak{q}_q}{\mathfrak{p}_q}\right|^2 = \left|\frac{\Gamma_{12}}{M_{12}}\right|\sin\phi_{12}\ +\mathcal{O}\left(\left|\frac{\Gamma_{12}}{M_{12}}\right|^2\sin^2\phi_{12} \right),
\end{equation}
which is small in the SM. The transitions in Eqs.~\eqref{eq: decay rate B to f} and \eqref{eq: decay rate Bbar to fbar} can occur directly and are usually referred to as ``right sign" decays. In contrast, the transitions in Eqs.~\eqref{eq: decay rate B to fbar} and \eqref{eq: decay rate Bbar to f} are called ``wrong sign'' decays as they can only proceed through oscillation.  

This modified time dependence can be physically understood as a manifestation of the Quantum Zeno effect \cite{Misra:1976by}. In vacuum, a non-zero width difference $\Delta \Gamma_q$ implies that the neutral meson is a coherent superposition of a short-lived and a long-lived mass eigenstates. At late times, the short-lived component naturally depletes, and the system becomes dominated by the long-lived configuration. In the plasma, however, each scattering event projects the wave function back onto a pure flavour state, which inherently contains an almost equal admixture of both mass eigenstates. This continuous resetting constantly regenerates the short-lived component, preventing the system from staying in the long-lived state, and therefore affecting the total decay rate. 

To determine the net baryon asymmetry, we integrate over the full decay time:
\begin{equation}\label{eq:ATIdef}
    \mathcal{A}_{\rm CP}^\text{TI}(B_q^0\to f)\equiv\int_0^\infty \left[ \Gamma(B^0_q(t)\to f)-\Gamma( B^0_q(t) \to \bar{f})\right]dt \ ,
\end{equation}
and equivalently for  $\mathcal{A}_{\rm CP}^\text{TI}(\bar{B}_q^0\to f)$.

Using the time-dependent rates given in Eqs.~\eqref{eq: decay rate B to f}-\eqref{eq: decay rate Bbar to f}, we find
\begin{align}
    \mathcal{A}_{\rm CP}^\text{TI}(B_q^0\to f){}&=(1+2x^\mathcal{S}_q)\frac{2(1-\tilde y_q^2)+a_q(\tilde x_q^2+\tilde y_q^2)}{2(1+\tilde x_q^2)(1-\tilde y_q^2)}\; \mathcal{B}(B_q^0\to f)|_{t=0},\label{eq:ATI1}\\
   \mathcal{A}^\text{TI}_{\rm CP}(\bar B_q^0\to f){}&=(1+2x^\mathcal{S}_q) \frac{-2(1-\tilde y_q^2)+a_q(2+\tilde x_q^2-\tilde y_q^2)}{2(1-a_q)(1+\tilde x_q^2)(1-\tilde y_q^2)}\;\mathcal{B}(\bar{B}_q^0\to f)|_{t=0} \label{eq:ATI2} \ ,
    \end{align} 
where we defined the effective mixing variables as 
\begin{equation}
    \tilde x_q^2\equiv x_q^2 + 2\frac{\Gamma_{eB}}{\Gamma_q}\, , \,\quad\tilde y_q\equiv y_q^2 - 2\frac{\Gamma_{eB}}{\Gamma_q},\quad x_q^\mathcal{S}=\frac{\Gamma_{eB}}{\Gamma_q},
\end{equation} in terms of the standard mixing variables $x_q\equiv\frac{\Delta M_q}{\Gamma_q}$ and $y_q\equiv\frac{\Delta \Gamma_q}{2\Gamma_q}$. Numerically, there is a large hierarchy between the $B_d^0$ and $B_s^0$ mass differences, where $x_d= 0.7697 \pm 0.0035$ and $x_s = 26.99 \pm 0.09$ \cite{ParticleDataGroup:2024cfk}, respectively. We note that $y_d\ll1$ and can be safely neglected. However, for the $B_s^0$ system, $y_s$ is sizeable and we have $y_s= 0.063 \pm 0.004$ \cite{ParticleDataGroup:2024cfk}. We stress again that the effective mixing parameters $\tilde x_q$ and $\tilde y_q$ depend on $\Gamma_{eB}$ and, therefore, on the plasma temperature.

The $B_q^0\to f$ branching ratio at $t=0$ is defined as \cite{DeBruyn:2012wj}, 
\begin{equation}
    \mathcal{B}(B^0_q\to f)|_{t=0}\equiv\frac{\mathcal{N}_f |A_f|^2}{\Gamma_q} ,
\end{equation}
and equivalently for $\mathcal{B}(\bar{B}^0_q\to f)|_{t=0}$. We assume there is no direct CP violation in the $B\to f$ decay, such that $A_f= \bar{A}_{\bar{f}}$ and $\mathcal{B}(\bar{B}^0_q\to f)|_{t=0} = \mathcal{B}({B}^0_q\to f)|_{t=0}$.

We are interested in the combined net asymmetry irrespective of whether the neutral meson is produced initially as a $B_q^0$ or $\bar B_q^0$. This asymmetry is defined as 
\begin{equation}
      \mathcal{A}^\text{TI}_{\rm CP}(\langle B_q^0\to f\rangle) \equiv   \mathcal{A}^\text{TI}_{\rm CP}( B_q^0\to f) +   \mathcal{A}^\text{TI}_{\rm CP}(\bar{B}_q^0\to f) \ .
\end{equation}

Adding Eqs.~\eqref{eq:ATI1} and \eqref{eq:ATI2}, we find
\begin{align}
    \mathcal{A}_\mathrm{CP}^\text{TI}(\langle B_q^0\to f\rangle) {}&=\frac{1}{2}(1+2x^\mathcal{S}_q)    \frac{a_q(-2+a_q)(\tilde x_q^2 + \tilde y_q^2)}{(-1+a_q)(1+\tilde x_q^2)(1-\tilde y_q^2)} \; \mathcal{B}(B_q^0\to f)|_{t=0} \ , \\
    {}& = (1+2x^\mathcal{S}_q)\frac{\tilde x_q^2+\tilde y_q^2}{(1+\tilde x_q^2)(1-\tilde y_q^2)}a_q\, \mathcal{B}(B_q^0\to f)|_{t=0}+\mathcal{O}(a_q^2) \ ,
    \label{eq: neutralasym}
\end{align}
where we expanded in the small parameter $a_q$.

This net baryon asymmetry $\mathcal{A}_\mathrm{CP}^\mathrm{TI}(\langle B_q^0\to f\rangle)$ is exactly the $\Gamma_{eB}$ dependent baryon asymmetry source $\Delta_{B_q^0}$ in Eq.~\eqref{eq: evolution individual baryon asymmetries}: 
\begin{equation}\label{eq:DeltaB0}
    \Delta_{B_q^0}= \mathcal{A}_\mathrm{CP}^\text{TI}(\langle B_q^0\to f\rangle) \ .
\end{equation}

\subsection{Hybrid $\boldsymbol{B_c}$ Mesogenesis} \label{sec:intro_hybmeso}
Finally, we discuss the baryon asymmetry source $\Delta_{B_c^+\to B_q^0}$. As discussed, in the hybrid $B_c$ mesogenesis mechanism, the baryon asymmetry receives contributions from direct CP violation in the primary decay $B_c^+ \to B_q^0$, and from mixing-induced CP violation arising from the subsequent $B_q^0 - \bar B_q^0$ oscillations.

We begin by assuming that equal numbers of $B_c^+$ and $B_c^-$ are produced from the $\Phi$ decay. In contrast to the neutral case, this scenario can involve an additional CP-violating decay step if $\,\Gamma(B_c^+\to B^0_qM^+)\neq \Gamma(B_c^-\to \bar B_q^0M^-)$, before the neutral-meson oscillations occur.

In terms of the time-integrated asymmetry $\mathcal{A}_{\rm CP}^{\rm TI}$ defined in Eq.~\eqref{eq:ATIdef}, the net baryon asymmetry is given by 
\begin{equation}
\begin{aligned}
    \Delta_{B_c^+ \to B_q^0}=\sum_{M^\pm}&\Big\{\Gamma(B_c^+ \to B^0_q M^+) \mathcal{A}_{\rm CP}^\mathrm{TI}(B_q^0\to f)  \\
    +& \Gamma(B_c^- \to \bar B^0_q M^-)\mathcal{A}_{\rm CP}^\mathrm{TI}(\bar{B}_q^0\to f) \Big\}\,.
    \end{aligned}
\end{equation}

In terms of observables, we then write
\begin{equation}\label{eq:DeltaBcneutral}
\begin{aligned}
    \Delta_{B_c^+\to B_q^0}=&\sum_{M^+}\,\mathcal{B}(B_c^+\to B_q^0M^+)\Bigl[ \mathcal{A}_{\rm CP}^{\rm TI}(\langle B_q^0 \to f \rangle )\\  &+ \mathcal{A}_\mathrm{CP}^\mathrm{dir}(B_c^+ \to B_q^0 M^+) \left( \mathcal{A}_{\rm CP}^\mathrm{TI}(B_q^0\to f) - \mathcal{A}_{\rm CP}^\mathrm{TI}(\bar B_q^0\to f) \right) \Bigr] \ ,
    \end{aligned}
\end{equation}
where the direct CP asymmetry is defined in Eq.~\eqref{eq:Adir} and the CP-averaged branching ratio in Eq.~\eqref{eq:Brdef}. 

Interestingly, in the difference $\mathcal{A}_{\rm CP}^\mathrm{TI}(B_q^0\to f) - \mathcal{A}_{\rm CP}^\mathrm{TI}(\bar B_q^0\to f)$ the leading contribution is independent of $a_q$.

\section{\boldmath Non-leptonic $B_c^+ \to BM$ decays in the SM} \label{sec:BcinSM}

\subsection{\boldmath Decay structure and observables} \label{sec:Bc_decstructure}
The $B_c^+ \to B$ transition proceeds through the weak decay of the $c$ quark, while the heavy $b$ quark only acts as a spectator. This stands in contrast to decays of the lighter $B$ mesons, where the $b$ quark is always involved in the weak transition. In this respect, the CKM structure of the $B_c^+ \to BM$ decays, where $B=\{B^+,B_d^0,B_s^0\}$ and $M$ is a light pseudoscalar meson, $M=P\in\{\pi,K,\eta,\eta'\}$ or a light vector meson, $M=V\in\{\rho,K^*,\omega\}$, is the same as that of two-body $D\to M_1M_2$ decays. As such, we distinguish the $B_c^+\to BM$ decays by their leading CKM structure: Cabibbo-favoured (CF) decays proportional to $V_{cs}^*V_{ud} \sim 1$, singly Cabibbo-suppressed (SCS) decays where the leading term is proportional to $V_{cs}^* V_{us}$ or $V_{cd}^*V_{ud}$ which are proportional to $\lambda \equiv |V_{us}|\sim 0.22$, and doubly Cabibbo-suppressed (DCS) decays proportional to $V_{cd}^*V_{us} \sim \lambda^2$.  

\begin{table}[t]
    \centering
    \renewcommand{\arraystretch}{1.25}
    \scalebox{0.85}{
        \begin{tabular}{lccccccc}
             \textbf{SCS Decay} & $T$ & $C$ & $P_{sd}$ & $P_{sd}^S$ & $P_{EW,sd}$ & $P_{EW,sd}^C$ \\
            \midrule
            \hline
             $B_c^+\to B_d^0\pi^+$ & $1$ & $0$ & $-1 $ &$0$ & $0$ & $-c_d $ \\[3mm]
             $B_c^+\to B^+\pi^0$ & $0$ & $-\frac{1}{\sqrt 2}$  & $-\frac{1}{\sqrt 2}$& $0$ & $-\frac{1}{\sqrt 2}(c_u-c_d)$ & $-\frac{1}{\sqrt{2}}c_u$ \\[3mm]
             $B_c^+\to B^+\eta_q$ & $0$ & $\frac{1}{\sqrt{2}}$  & $-\frac{1}{\sqrt 2}$ & $-\sqrt{2}$ & $-\frac{1}{\sqrt{2}}(c_u+c_d)$ & $-\frac{1}{\sqrt{2}}c_u$ \\[3mm]
             \hline
             $B_c^+\to B_s^0K^+$ & $1$ & $0$ & $1 $ & $0$& $0$ & $c_s $ \\[3mm]
             $B_c^+\to B^+\eta_s$ & $0$ & $1$ &  $0$ & $1$  & $c_s $ & $0$ \\[3mm]
\hline \hline
        \end{tabular}
    }
    \caption{Topological amplitude decomposition for the CKM leading amplitude of the SCS $B_c^+ \to B P$ decay modes proportional to $\lambda_d$ (upper) and $\lambda_s$ (lower). $c_q$ represents the electric charge of the quark $q$.} 
    \label{tab:Bc to B topologies}
\end{table} 

The CF and DCS decays proceed through a single combination of CKM coefficients. As such, there is no direct CP violation in these decays. In contrast, SCS decays involve multiple interfering CKM combinations due to richer amplitude structures, giving rise to potential CP violation. 

The SCS decay amplitude can be written as
\begin{equation}
     \mathcal{A}(B_c^+ \to B M)=i\frac{G_F}{\sqrt{2}} \left[\lambda_D \mathcal{T}_i + \lambda_b \mathcal{P}_i \right] \ ,
     \label{eq: Bc amplitude}
 \end{equation}
with $D=d,s$ depending on the underlying $c\to d\bar{d}u$ and $c\to s\bar{s}u$ transitions, respectively, and $\lambda_q \equiv V_{cq}^*V_{uq}$.

The CKM leading (tree) amplitude is proportional to $\lambda_s = -\lambda_d + \mathcal{O}(\lambda^5) \sim \lambda$. The $\mathcal{P}_i$ terms are suppressed by $\lambda_b/\lambda_D \sim \lambda^4 \sim 10^{-3}$ with respect to the CKM leading $\mathcal{T}_i$ terms, and can therefore be safely neglected when considering the branching ratio defined in Eq.~\eqref{eq:Brdef}. 

In order to give predictions for the branching ratios, we write the leading CKM amplitude $\mathcal{T}_i$ in terms of the topological amplitudes: 
colour-allowed tree $T$, colour-suppressed tree $C$, QCD penguin $P$, singlet-QCD penguin $P^S$, colour-allowed EW-penguin $P_{EW}$ and colour-suppressed EW-penguin $P_{EW}^C$. The contribution of the different topologies to $\mathcal{T}_i$ is given in Table~\ref{tab:Bc to B topologies}. Here $P_{sd} \equiv P_s - P_d$, where $P_q$ represents a penguin topology with an internal $q$-quark. An analogous definition applies to the electroweak (EW) penguin contributions, which are proportional to the electric charge $c_q$ of the $q$--$\bar{q}$ pair generated from the $\gamma$.  Both the $P_{sd}^{(S)}$ and $P_{EW, sd}^{(C)}$ contributions vanish in the massless light-quark limit and are expected to be suppressed. The leading topologies are the $T$ and $C$ topologies, depicted in Fig.~\ref{fig:T_C_diags}. The QCD penguin and EW penguin diagrams, with $d,s,b$ quarks in the loop are given in Fig.~\ref{fig: penguin diagrams}. 

For completeness, we list the CKM subleading $\mathcal{P}_i$ terms in Appendix~\ref{ap:CKM_sublead}, although we do not further study these terms in this work. In the current work, we only estimate the branching ratios of the $B_c\to BM$ decays, for which the $\mathcal{P}_i$ terms can be neglected. However, these terms play a crucial role in the CP asymmetry, defined in Eq.~\eqref{eq:Adir}. We can write 
\begin{equation}\label{eq:adircharm}
    \mathcal{A}_{\rm CP}^{\rm dir} = -2\;{\rm Im}\;\frac{\lambda_b}{\lambda_{D}} \; {\rm Im}\frac{\mathcal{P}_i}{\mathcal{T}_i} \sim \mathcal{O}(10^{-3}) \; {\rm Im}\frac{\mathcal{P}_i}{\mathcal{T}_i}  \ ,
\end{equation}
where we neglected further subleading corrections. The numerical estimate comes from the CKM suppression. 
The relative strong-phase difference between the CKM leading (tree ) $\mathcal{T}_i$ terms and the CKM subleading (penguin) $\mathcal{P}_i$ terms is challenging to calculate already for $B$ decays, and even more so for decays of the charm quark. In this work, we do not further attempt a SM determination of the CP asymmetries.
Instead, our goal is to identify the range of asymmetry values that would allow these channels to contribute effectively to baryogenesis as discussed below. 
Decays with vector mesons involve the same topological structure as their pseudoscalar counterparts and can be obtained from Table~\ref{tab:Bc to B topologies} through the replacements: $\pi^{\pm,0} \to \rho^{\pm,0}$, $K^{\pm,0} \to (K^*)^{\pm,0}$ and $\eta_{q,s} \to \omega_{q,s}$.  In Table~\ref{tab:Bc to B topologies}, we expressed the amplitudes in terms of the flavour eigenstates $\eta_q$ and $\eta_s$. Modes involving $\eta$ and $\eta'$ mesons are generally more complex due to the mixing between octet and singlet states. We detail in Appendix~\ref{ap:eta} how we relate the flavour eigenstates to the physical $B_c^+\to B^+ \eta$ and $B_c^+\to B^+\eta'$ amplitudes. 

\begin{figure}[t]
\centering
\begin{subfigure}[b]{0.43\textwidth}
\begin{tikzpicture}
  \begin{feynman}
    \vertex (a) ;
    \vertex at ($(a) + (2cm, 0 cm)$) (b);
    \vertex at ($(b) + (2.6cm, -1 cm)$) (c);
    \vertex at ($(a) + (0cm, -1cm)$)(d) ;
    \vertex at ($(d) + (1.7cm, 0cm)$)(e) ;
    \vertex at ($(e) + (2.9 cm, -1.04cm)$) (f);
    \vertex at ($(b) + (1.1 cm, 1.1cm)$) (g);
    \vertex at ($(g) + (1.5 cm, 0.5cm)$) (h);
    \vertex at ($(g) + (1.5 cm, -0.5cm)$) (i);
    \node[draw,ellipse,minimum width=1cm,minimum height=0.5cm, rotate=90,fill=gray!30,
  fill opacity=0.4] (Bc) at ($(a)!0.5!(d)$) {};
  \node[draw,ellipse,minimum width=1cm,minimum height=0.5cm, rotate=90,fill=gray!30,
  fill opacity=0.4] (B0) at ($(c)!0.5!(f)$) {};
  \node[draw,ellipse,minimum width=1cm,minimum height=0.5cm, rotate=90,fill=gray!30,
  fill opacity=0.4] (pip) at ($(h)!0.5!(i)$) {};
  \node[anchor=east, xshift=-0.25cm] at (Bc.center) {$B_c^+$};
  \node[anchor=west, xshift=0.25cm] at (B0.center) {$B_d^0$};
  \node[anchor=west, xshift=0.25cm] at (pip.center) {$\pi^+$};
    
    \diagram* {
      (a) -- [fermion, edge label=\(c\)] (b);
      (b) -- [fermion, edge label=\( d\)] (c),
      (e) -- [fermion, edge label=\(\bar b\)] (d);
      (f) -- [fermion, edge label=\(\bar b\)] (e);
      (b) -- [photon, edge label=\(W\)] (g);
      (g) -- [fermion, edge label=\(u\)] (h);
      (i) -- [fermion, edge label=\(\bar d\)] (g);
    };
  \end{feynman}
\end{tikzpicture}
  \caption{colour-allowed tree topology, $T$}
\end{subfigure}
\begin{subfigure}[b]{0.43\textwidth}
\begin{tikzpicture}
  \begin{feynman}
    \vertex (a) ;
    \vertex at ($(a) + (2cm, 0 cm)$) (b);
    \vertex at ($(b) + (2.6cm, 1 cm)$) (c);
    \vertex at ($(a) + (0cm, -1cm)$)(d) ;
    \vertex at ($(d) + (2cm, 0cm)$)(e) ;
    \vertex at ($(e) + (2.6 cm, -1cm)$) (f);
    \vertex at ($(b) + (1.1 cm, -0.5cm)$) (g);
    \vertex at ($(g) + (1.5 cm, 0.5cm)$) (h);
    \vertex at ($(g) + (1.5 cm, -0.5cm)$) (i);
    \node[draw,ellipse,minimum width=1cm,minimum height=0.5cm, rotate=90,fill=gray!30,
  fill opacity=0.4] (Bc) at ($(a)!0.5!(d)$) {};
  \node[draw,ellipse,minimum width=1cm,minimum height=0.5cm, rotate=90,fill=gray!30,
  fill opacity=0.4] (Bp) at ($(i)!0.5!(f)$) {};
  \node[draw,ellipse,minimum width=1cm,minimum height=0.5cm, rotate=90,fill=gray!30,
  fill opacity=0.4] (pi0) at ($(c)!0.5!(h)$) {};
  \node[anchor=east, xshift=-0.25cm] at (Bc.center) {$B_c^+$};
  \node[anchor=west, xshift=0.25cm] at (Bp.center) {$B^+$};
  \node[anchor=west, xshift=0.25cm] at (pi0.center) {$\pi^0$};
    
    \diagram* {
      (a) -- [fermion, edge label=\(c\)] (b);
      (b) -- [fermion, edge label=\( d\)] (c),
      (e) -- [fermion, edge label=\(\bar b\)] (d);
      (f) -- [fermion, edge label=\(\bar b\)] (e);
      (b) -- [photon, edge label=\(W\)] (g);
      (h) -- [fermion, edge label'=\(\bar d\)] (g);
      (g) -- [fermion, edge label'=\(u\)] (i);
    };
  \end{feynman}
\end{tikzpicture}
\caption{colour-suppressed tree topology, $C$}
\end{subfigure}
\caption{Leading topological diagrams for $B_c^+ \to B_d^0 \pi^+$ and $B_c^+ \to B^+ \pi^0$.}
\label{fig:T_C_diags}

\end{figure}
The CF decays $B_c^+\to B_s^0\pi^+$ and $B_c^+\to B^+\bar K^0$, as well as their vector counterparts, involve four different quark flavours, therefore these mode are described by a single tree topology: the colour-allowed tree $T$ and colour-suppressed tree $C$, respectively. Similarly, the DCS $B_c^+\to B_d^0 K^+$ decay is only mediated by $T$, and the $B_c^+\to B^+ K^0$ only by $C$.

\subsection{Factorization of \boldmath $B_c^+ \to BM$ decays} \label{sec:factinBc}
The calculation of the $B_c^+\to B M$ decay rate is challenging, as is the case for any purely hadronic $B$ decay, because of non-perturbative QCD effects. The light $B\to M_1 M_2$ decays have been studied extensively using QCD Factorization \cite{Beneke:1999br, Beneke:2000ry, Beneke:2003zv} and/or using $SU(3)_F$ flavour symmetry (see e.g.~\cite{Fleischer:2022rkm, Huber:2021cgk, Berthiaume:2023kmp, Bhattacharya:2025wcq, BurgosMarcos:2025xja}). 
However, the factorization in $B_c^+\to BM$ decays may be even more challenging as the $b$ acts as a spectator, while the charm quark decays through the weak interaction. Therefore, non-factorizable corrections would be expected to be suppressed by powers of $\frac{\Lambda_{QCD}}{m_c}$ only, similar as in $D$ decays. Compared to the lighter $B$ decays, where such terms scale as $\frac{\Lambda_{QCD}}{m_b}$, these corrections are potentially more significant. In addition, the scale of the perturbative corrections is also considerably lower, making these effects more significant. On the other hand, for annihilation modes, as discussed in Ref.~\cite{Descotes-Genon:2009eui,Liu:2009qa,Wang:2016qli}, the mass of the charm quark regulates the typical end-point divergences that arise in models for the weak annihilation modes \cite{Beneke:2003zv}. In addition, spectator scattering effects in the light $B\to M_1M_2$ decays suffer from end-point divergences due to an enhancement of the twist-3 contributions of the light-meson LCDAs. Here, the momentum scaling of the quarks is more similar to $B\to D M$ decays, where based on power-counting arguments these terms are power suppressed. A full discussion of the theoretical treatment of $B_c^+\to B M$ decays lies beyond the scope of this work.

In the following, we use two approaches to estimate the $B_c^+\to BM$ branching ratios. For both approaches, we first factorize the CKM leading $\mathcal{T}_i$ amplitude which is mediated by the weak effective Hamiltonian (see e.g.~\cite{Buchalla:1995vs}). Schematically, we decompose the relevant matrix elements in terms of form factors and decay constants:
\begin{equation}
    \langle BM|\mathcal{O}_2|B_c^+\rangle\simeq \langle B|\bar q \gamma_\mu (1-\gamma_5) c|B_c^+\rangle \times  \langle M|\bar u\gamma^\mu(1-\gamma_5)q |0\rangle \ ,
    \label{eq:fact_formula}
\end{equation}
where we illustrate factorization using the current--current operator $\mathcal{O}_2=(\bar{q}^i\gamma_\mu(1-\gamma_5)c^i)(\bar{u}^j\gamma^\mu(1-\gamma_5)q^j)$, which mediates the $c \to u q  \bar q $ colour-allowed tree topology. Here, the first term is the $B_c^+ \to B$ form factor defined as 
\begin{equation}
    \langle B(p_B)|\bar q_2 \gamma_\mu c|B_c^+(p_{B_c^+})\rangle=F_+^{B_c^+ \to B}(k^2)(p_{B_c^+}+p_B)_\mu + k_\mu \frac{m_{B_c^+}^2-m_{B}^2}{k^2}\left[ F_0^{B_c^+\to B}(k^2)-F_+^{B_c^+\to B}(k^2)\right],
    \label{eq:form_factor}
\end{equation}
with $k=p_{B_c^+}-p_B$. The second term is the decay constant, 
\begin{equation}
    \langle P(k)|\bar u\gamma^\mu\gamma_5q_1|0\rangle=-i f_P k^\mu  \ ,\hspace{1cm}
    \langle V(k)|\bar u\gamma^\mu q_1|0\rangle=-i f_V m_V\epsilon^{*\mu} \ ,
    \label{eq:decayconst}
\end{equation}
respectively for a pseudoscalar, $M=P$, or a vector meson, $M=V$.

We can then parametrize the CKM leading $\mathcal{T}_i$ topologies as 
\begin{equation}\label{eq:fact}
    T\equiv A_{BM}^{Bc^+}\, \alpha_1(BM)  \ , \quad\quad 
    C\equiv  A_{BM}^{Bc^+}\, \alpha_2(BM)  \ ,
\end{equation}
and similarly for the suppressed penguin-type contributions. 
The $A_{BM}^{B_c^+}$ parameter is defined as
\begin{equation}
A_{BP}^{B_c^+}\equiv F_0^{B_c^+ \to B}(m_P^2)f_P (m_{B_c^+}^2-m_B^2)\ , \hspace{1cm} A_{BV}^{B_c^+}\equiv -F_+^{B_c^+ \to B}(m_V^2)f_V m_V \epsilon^{*}\cdot(p_{B_c^+}+p_B) \ , 
\label{eq:fact_prefactor}
\end{equation}
for pseudoscalar and vector mesons, respectively. For decays involving $\eta$ and $\eta'$ mesons in the final state we take the decay constants for the physical mesons.

In the branching ratio, $|A_{BM}^{B_c^+}|^2$ appears summed over all possible polarizations. In the case of a vector meson in the final state, this allows for the replacement: 
\begin{equation}
    -m_V\epsilon^{*}\cdot(p_{B_c^+}+p_B) \to \sqrt{(m_{B_c^+}^2-m_B^2)^2-m_V^2(2m_{B_c^+}^2+2m_B^2-m_V^2)}.
    \label{eq: vector term}
\end{equation}
 We stress that for $B_c^+\to B M$ decays, the $m_V^2$ terms are numerically important. The small mass splitting between the $B_c$ and $B$ mesons implies that $(m_{B_c^+}^2 - m_B^2)$ is of the same order as $m_V^2$, and therefore these terms cannot be neglected.  

We note that the parametrization in Eq.~\eqref{eq:fact} is general. Known non-perturbative corrections are factored out through $A_{BM}^{B_c^+}$. The coefficients $\alpha_1$ and $\alpha_2$ depend on the Wilson coefficients of the current-current operators. In addition, they contain {\it factorizable} perturbative corrections\footnote{For the light $B\to M_1M_2$ decays these coefficients are known up to NNLO \cite{Beneke:2009ek, Bell:2009nk}} as well as {\it non-factorizable} corrections.

\subsubsection{Leading-Order Factorization Approach}
First, we only consider the leading topologies: the colour-allowed tree, $\alpha_1$, and the colour-suppressed tree, $\alpha_2$. To estimate these coefficients, we neglect all subleading and/or non-factorizable contributions as well as perturbative $\alpha_s$ corrections. 

From the effective Hamiltonian, we find, at leading order,
\begin{equation}
    \alpha_1=C_2+\frac{C_1}{N_c},
\end{equation}
where $N_c=3$ is the number of colours and $C_i$ the Wilson coefficients of the current--current operators, as defined in \cite{Buchalla:1995vs}, and evaluated at the scale set by the charm mass. Working at leading order also implies that $\alpha_1$ is the same for all decays considered. Therefore $SU(3)_F$-symmetry breaking effects only appear via $A_{BM}^{B_c^+}$, through form factor, decay constants and phase-space effects. 

Taking $\mu=m_c\approx 1.48  \,\mathrm{ GeV}$ and the values for the Wilson coefficients from \cite{King:2021xqp}\footnote{Note that in \cite{King:2021xqp} the operator $\mathcal{O}_1$ is the singlet operator.}, we obtain:
\begin{equation}
    \alpha_1 = 1.06 \ .
\end{equation} 
Although we do not assign uncertainties to the Wilson coefficients in our numerical analysis, they could be estimated through the usual variation of the charm-mass scale. This has only a minor effect on $C_2$ but a noticeably larger impact on $C_1$. 

The situation is more delicate for the colour-suppressed amplitude $\alpha_2$. Besides inheriting the stronger scale sensitivity of $C_1$, the leading-order combination $C_1 + \frac{C_2}{N_c}$ is subject to an almost accidental numerical cancellation \cite{Buras:1994ij, Silvestrini:1999vs}, implying that higher-order contributions may compete with, or even exceed, the nominal leading term. To avoid an artificially suppressed prediction, and given the structural similarity between the $\alpha_1$ and $\alpha_2$ topologies aside from colour factors, we adopt the approximation
\begin{equation}
    \alpha_2\approx \frac{\alpha_1}{N_c}.
\end{equation}

\subsubsection{\boldmath Connecting to hadronic $D$ decays} \label{sec:datadriven_approach}
As discussed, going beyond the leading-order factorization approach is theoretically challenging. It is therefore interesting to compare to experimental data in order to estimate the size of higher-order corrections and subleading topologies. Since (almost) no measurements are currently available for the $B_c$ decays, we use information from $D \to M_1M_2$ transitions. The topologies that contribute to these decays are listed in e.g.~\cite{Muller:2015lua}. The underlying dynamics of these two channels are closely related, differing essentially only in the flavour of the spectator quark. Interestingly, in terms of the topological decomposition, almost every $B_c^+\to BM$ decay has a fully equivalent $D\to M_1M_2$ decay\footnote{Small differences appear in subleading annihilation modes as discussed in Appendix~\ref{ap:DMM_decays}.}, which is found by replacing the spectator $b$ by a $u, d$ or $s$ quark. The charm partner modes of the relevant $B_c^+\to BM$ decays are given in Table~\ref{tab:exp_DtoMM} of Appendix~\ref{ap:DMM_decays}.

We start by factorizing the $D\to M_1M_2$ decays\footnote{Recently, the $D\to \pi K, KK, \pi\pi$ were studied in leading-order factorization \cite{Lenz:2023rlq,Fleischer:2025zhl}.}. The topologies can then be parametrized similar to Eq.~\eqref{eq:fact}, where now the colour-allowed and colour-suppressed amplitudes can be parametrized as
\begin{equation}\label{eq:factD}
    T_D\equiv A_{M_1M_2}^{D}\, \alpha_1^D(M_1M_2)  \ , \quad\quad 
    C_D\equiv  A_{M_1M_2}^{D}\, \alpha_2^D(M_1M_2)  \ ,
\end{equation}
respectively. We note that in our case, $M_1$ is always a pseudoscalar. In addition, we have
\begin{equation}
    A_{ M_1  P_2}^D\equiv m_D^2 F_0^{D\to  M_1}(0)f_{ P_2},
\end{equation}
for pseudoscalar final states, where we neglected quadratic mass terms. For vector mesons, the expression can be obtained from Eq.~\eqref{eq:fact_prefactor} with analogous replacements.

Assuming that we have $\alpha_i(BM) = \alpha_i^D(M_1M_2)$ for partner decays that share the same topological structure, i.e. neglecting spectator quark differences, the branching ratios can be related as
\begin{equation}
    \mathcal{B}(B_c^+\to BM)=\frac{\tau_{B_c^+}}{\tau_D}\frac{m_{D}}{m_{B_c^+}}\frac{\Phi\left(m_{B}/m_{B_c^+},m_{M}/m_{B_c^+}\right)}{\Phi\left(m_{M_1}/m_{D}, m_{ M_2}/m_{D}\right)}\left( \frac{A_{BM}^{B_c^+}}{A_{ M_1 M_2}^D}\right)^2\mathcal{B}(D\to  M_1 M_2) \ .
    \label{eq:BRDtoBc}
\end{equation}

For some modes, mainly DCS decays, the partner $D$-meson channel has not (yet) been measured. In these cases, the relevant branching ratios are adapted from other $B_c^+ \to BM$ decays, as detailed in Appendix \ref{ap:DMM_decays}. 

Finally, for $B_c^+ \to B^+ \eta^{(')}$ decays, changing the spectator to the light $u,d,s$ quarks always corresponds to a $D$ mode that has additional topologies with respect to the original $B_c$ decay. As a result, there are no partner $D$ modes for these decays and we cannot follow the above approach. For these decays, we thus restrict ourselves to the leading-order approach.

\subsection{Branching Ratio predictions} \label{sec:BRpred}
In this section we give branching ratio predictions using both the leading-order factorization approach and the data-driven approach. 

For our numerical analysis, we use the CKM elements from the CKMFitter Collaboration \cite{Charles:2004jd}, while masses and lifetimes are taken from PDG \cite{ParticleDataGroup:2024cfk}. Decay constants and form factors are listed in Table~\ref{tab: dec constants and ffs}.  We take the form factor at the scale $q^2= m_M^2$, although these quadratic light-meson mass effects are subleading compared to those in Eq.~\eqref{eq: vector term}. We adopt a simple parametrization based on vector-meson dominance:
\begin{equation}
    F_{0,+}^{B_c^+ \to B}(m_M^2) \simeq \frac{F_{0,+}^{B_c^+\to B}(0)}{1-m_M^2/m_\mathrm{res}^2},
\end{equation}
where $m_\mathrm{res}$ denotes the mass of the $c\bar p$ resonance associated with the $c \to p$ transition: scalar meson resonances ($D_{s0}^*$ and $D_{0}^*$) for $F_0^{B_c^+ \to B}$ and vector ones ($D_{s}^*$ and $D^*$) for $F_+^{B_c^+ \to B}$. 
\begin{table}[t]
   \centering
   \renewcommand{\arraystretch}{1.25}
    \setlength{\tabcolsep}{4pt}

    {\tabcolsep=1.38cm\begin{tabular}{cccc}
    \multicolumn{4}{c}{Pseudoscalar and Vector meson decay constants [MeV]} \\
    \hline \hline
    $f_\pi$ & $f_K$ & $f_\eta$ & $f_{\eta'}$\\
    \hline
    $130.2$ \cite{FlavourLatticeAveragingGroupFLAG:2024oxs} & $155.7$ \cite{FlavourLatticeAveragingGroupFLAG:2024oxs} & $117.6$ \cite{Bali:2021qem} & $49$ \cite{Bali:2021qem}\\
    \hline \hline
    \end{tabular}}
    {\tabcolsep=2.07cm\begin{tabular}{ccc}
$f_\rho$ & $f_{K^*}$  & $f_\omega$  \\
\hline
215 \cite{Arifi:2025olq} & 218 \cite{Beneke:2003zv} & 187 \cite{Beneke:2003zv}\\
\hline \hline\\
\end{tabular}}
    {\tabcolsep=0.8cm\begin{tabular}{ccccc}
    \multicolumn{5}{c}{Form factors for pseudoscalar mesons (at $q^2=0$)} \\
    \hline \hline
      $F_0^{D\to\pi}$ & $F_0^{D\to K}$ & $F_0^{D_s\to K}$ & $F_0^{B_c^+\to B^0}$ & $F_0^{B_c^+\to B_s^0}$\\
      \hline
     $0.630$\cite{FlavourLatticeAveragingGroupFLAG:2024oxs} & $0.743$\cite{FlavourLatticeAveragingGroupFLAG:2024oxs} & $0.65$ \cite{FermilabLattice:2022gku}& $0.555$ \cite{Cooper:2020wnj} & $0.621$ \cite{Cooper:2020wnj}\\
     \hline \hline
    \end{tabular}}
    \caption{Decay constants and form factors inputs.}
    \label{tab: dec constants and ffs}
\end{table}

Our results are given in Table~\ref{tab:BctoB_BRs}, expressed in percentage, to highlight the difference in order of magnitude between CF and DCS decays. In general, decays involving a light vector meson are enhanced relative to their pseudoscalar counterparts, although in some modes this effect is partially compensated by phase-space suppression. Phase space plays a non-negligible role in $B_c^+\to BM$, with the largest suppression occurring in $B_c^+ \to B_s^0 K^{*,+}$ through $\Phi(m_{B_s^0}/m_{B_c^+}, m_{K^{*,+}}/{m_{B_c^+}})\sim 0.05$. Additionally, we observe a strong suppression in the $\eta'$ mode due to, among other effects, partial cancellation in the amplitude mixing (see Appendix~\ref{ap:eta}).

Overall, the two methods yield roughly consistent results, with a few modes that differ by up to an order of magnitude. We do not quote uncertainties on our estimates of the branching ratios, which are obtained using factorization assumptions. For charm decays, these assumptions rely on an expansion in $\alpha_s(m_c) \sim 0.3$ and $\Lambda_{\rm QCD}/m_c \sim 0.3$ at the amplitude level, and could thus be subject to large uncertainties. Our results are estimates, to be verified with experimental measurements. Once data is available, it would be interesting to also further study these decays using additional flavour symmetries assumptions. We can already compare with the only measured mode in this class, $B_c^+\to B_s^0 \pi^+$. The experimental value in Eq.~\eqref{eq:Bcmeas}, which also depends on the value for the fragmentation fraction $f_c$, from \cite{LHCb:2013xlg, LHCb:2019tea} has $\sim 30\%$ uncertainty and agrees with our results at the $(2-3)\sigma$ level. Therefore, further experimental updates of this mode would be interesting to further test factorization in these modes. 

At the same time, the agreement between the two approaches is of interest. In the data-driven approach, it is assumed that non-factorizable corrections and subleading contributions in $B_c^+ \to BM$ decays are equivalent to those in the corresponding $D\to M_1M_2$ mode. Since these specific effects are neglected in the LO-Factorization approach, the difference between the two can be viewed as an estimate of the uncertainty associated with these corrections. In the following, we use these estimates to test the viability of $B_c$ mesogenesis. 
\begin{table}[t]
\centering
    \renewcommand{\arraystretch}{1.5}
    \scalebox{0.9}{
    \begin{tabular}{ll@{\hskip 15mm}ll}

 $B_c^+ \to BP$ decay & $\mathcal{B}^{\rm fac} \left[\mathcal{B}^{\rm dd}\right]$ $ \mathrm{in}\,\%$ & $B_c^+ \to BV$ decay & $\mathcal{B}^{\rm fac} \left[\mathcal{B}^{\rm dd}\right]$ $\mathrm{in}\,\%$\\
\hline
\hline
$B_c^+ \to B^+ \bar K^0$ & $0.51  \; [1.9] $  & 
$B_c^+ \to B^+ \bar K^{*,0}$ & $0.14 \; [1.4]$ \\
$B_c^+ \to B^0_s \pi^+$ & $ 3.3 \hspace{3mm} [2.7]$ & 
$B_c^+ \to B^0_s \rho^+$ & $1.6 \hspace{3mm} [1.8]$\\
\hline
$B_c^+ \to B^+ \pi^0 $ & $ 1.0 \hspace{3mm} [4.8] \times 10^{-2}$& $B_c^+ \to B^+ \rho^0 $ & $ 0.88 \; [5.8] \times 10^{-2}$\\
$B_c^+ \to B^+ \eta $ & $2.2 \hspace{3mm} [{\rm x}]\hspace{1.5mm}\times 10^{-2}  $ & $B_c^+ \to B^+ \omega $ & $ 0.67 \;[2.6] \times 10^{-2}$\\
$B_c^+ \to B^+ \eta' $ & $ 8.7 \hspace{3mm} [\rm{x}]\hspace{1.5mm}\times 10^{-5}$ & $- $ & $-$\\
$B_c^+ \to B^0_d \pi^+ $ & $ 0.18 \; [0.13]$ &
$B_c^+ \to B^0_d \rho^+ $ & $0.16\; [0.38]$\\
$B_c^+ \to B_s^0 K^+ $ & $ 0.23\; [0.17] $ & 
$B_c^+ \to B_s^0 K^{*,+} $ & $ 5.0 \hspace{3mm}[3.6^\star] \times 10^{-3}$\\
\hline
$B_c^+ \to B^+ K^0 $ & $ 1.5 \hspace{3mm}[5.6^{\star}]  \times 10^{-3}$ &
$B_c^+ \to B^+ K^{*,0} $ & $ 0.39 \;[4.0^{\star}] \times 10^{-3}$\\
$B_c^+ \to B^0_d K^+ $ & $ 1.3\hspace{3mm}[1.1^{\star}] \times 10^{-2}$ &
$B_c^+ \to B^0_d K^{*,+} $ & $ 2.4 \hspace{3mm} [3.1^\star] \times 10^{-3}$\\
\hline
\hline
\end{tabular}
}
\caption{Estimations for the branching ratios for $B_c^+ \to BM$ decays in $\%$ using the leading-order factorization approach $\mathcal{B}^\mathrm{fac}$, and the data-driven approach $\mathcal{B}^\mathrm{dd}$. Entries marked with ``x'' and ``$\star$'' correspond to modes for which a direct comparison with $D \to M_1M_2$ decays is not possible. For the latter, a comparison is made with other $B_c$ modes, as described in Appendix~\ref{ap:DMM_decays}. }

\label{tab:BctoB_BRs}
\end{table}

\section{Estimates for individual contributions to the baryon asymmetry} \label{sec:numerical_results}

\subsection{\boldmath Neutral $B$ Mesogenesis }\label{sec: estimates from neutral B mesogenesis}

The predictions for the baryon asymmetry in neutral $B$ mesogenesis depend on several SM inputs. While parameters such as the mass- and width-difference ratios $x_q$ and $y_q$ are known with good precision (see Sec.~\ref{sec:intro_neutmeso}), others are much less constrained. This is notably the case for $a_q$ defined in Eq.~\eqref{eq:aqdef}, which can be extracted from the measurements of the semileptonic asymmetry. In the following, we therefore refer to $a_q$ as the semileptonic asymmetry.

The experimental measurement of the semileptonic asymmetry reads
\cite{HeavyFlavorAveragingGroupHFLAV:2024ctg}
    \begin{equation}\label{eq:aexp}
        a_d^\text{exp}=(-2.1\pm 1.7)\times 10^{–3}, \hspace{1.5cm} a_s^\text{exp}=(-0.6 \pm 2.8)\times 10^{-3} ,
    \end{equation}
which has a sizeable uncertainty. On the other hand, the theoretical calculations are much more precise, where recently NNLO QCD corrections have been obtained \cite{Nierste:2025muk}:
\begin{equation}\label{eq:atheo}
    a_d^\mathrm{theo}=(-5.19\pm 0.30)\times 10^{-4} ,\hspace{1.5cm} a_s^\mathrm{theo}= (2.27\pm 0.13)\times 10^{-5},
\end{equation}
The large difference in theoretical and experimental precision leaves room for possible new physics effects (see \cite{Miro:2024fid} for a recent discussion in connection to $B$ mesogenesis). Since the viability of $B$ mesogenesis models involving neutral $B_q^0$ mesons is proportional to $a_q$, a nonzero value is essential for baryon asymmetry generation. In the following, we use both the theoretical and experimental determinations. 

Using Eq.~\eqref{eq:DeltaB0} and setting the reheating temperature to $T_R= 25 \;\mathrm{MeV}$, we find 
\begin{equation}
    \begin{aligned}
    Y_\mathfrak{B}^{B_d^0} &= 1.3 \times 10^{-6}\;a_d\;\mathcal{B}(B\to \mathfrak{B}\xi\phi) \ ,  \\
    Y_\mathfrak{B}^{B_s^0} &= 2.9\times 10^{-4} \; a_s\; \mathcal{B}(B\to \mathfrak{B}\xi\phi)  \ .\\
    \end{aligned}
    \label{eq: YB neutral general}
\end{equation}
where the numerical uncertainty arising from the mixing parameters $x_q$ and $y_q$ is not explicitly quoted as it remains below the percent level. These numerical results highlight the critical role of the semileptonic asymmetries $a_q$ in shaping the final baryon asymmetry.

To see how this compares to the required total baryon asymmetry, we set $\mathcal{B}(B\to \mathfrak{B}\xi\phi)\simeq 10^{-3}$ and fix the mass of the $\Phi$ field to $m_\Phi = 100 \,\mathrm{GeV}$. Note that while the $\Delta_X$ terms in Eq.~\eqref{eq: evolution individual baryon asymmetries} are independent of the chosen $m_\Phi$, the asymmetries $Y_\mathfrak{B}^X$ depend inversely on this parameter. Considering both the experimental and theoretical determinations of $a_q$ we find
\begin{align}\label{eq:res0}
  Y_\mathfrak{B}^{B_d^0,\rm exp} &= (-2.7 \pm 2.2) \times 10^{-12}\, \quad\quad   Y_\mathfrak{B}^{B_s^0,\rm exp} = (-1.7 \pm 8.0) \times 10^{-10} \ , \\
    Y_\mathfrak{B}^{B_d^0,\rm theo} &= (-6.7 \pm 0.4) \times 10^{-13}\, \quad\quad   Y_\mathfrak{B}^{B_s^0,\rm theo} = (6.5 \pm 0.4) \times 10^{-12} \ , \label{eq:res01}
\end{align}
where the quoted uncertainties arise from $a_q$. 

The contributions of the $B_d^0$ and $B_s^0$ modes to the total observed baryon asymmetry are obtained by multiplying with their respective fragmentation fractions following Eq.~\eqref{eq: individual Y's}. 
We recall that the observed baryon asymmetry is $Y_\mathfrak{B}\simeq 8.72 \times 10^{-11}$. 

At the 1$\sigma$ level, the positive contribution from the $B_s^0$ channel compensates the negative impact of the $B_d^0$ meson, for the typical fragmentation fractions $f_s/f_d\sim 0.24$ observed at colliders \cite{HeavyFlavorAveragingGroupHFLAV:2024ctg, LHCb:2021qbv}. Negative results of $Y_\mathfrak{B}$ imply an anti-matter dominated universe, in clear conflict with observations. Physically, the dominance of the $B_s^0$ channel is due, in part, to its higher resilience against electron rescattering in the early universe, since its rapid oscillations protect it from thermal decoherence. 

Overall, the current experimental uncertainties still allow the neutral $B$ mechanism to range from being strongly constrained to remaining a viable source of the baryon asymmetry at the $1.4\sigma$ level. Updated measurements will therefore be crucial to clarify the viability of neutral $B$ mesogenesis. On the other hand, considering $a_q^{\rm theo}$ imposes significantly stronger constraints on $Y_\mathfrak{B}$. In this case, the $B_s^0$ contribution still dominates over the $B_d^0$ one by one order of magnitude. However, the $a_q^\mathrm{theo}$ are overall significantly smaller in magnitude, therefore even under the assumption of maximal $B_s^0$ meson production, that is if only $B_s^0$ mesons would be produced, $Y_{\mathfrak{B}}$ would not be large enough to explain the observed baryon asymmetry.

Finally, we note that, although experimental determinations of $a_q$ offer a much wider window for accommodating the observed $Y_\mathfrak{B}$ compared to the theoretical predictions, achieving values of $a_q$ that deviate significantly from the theoretical calculations remains highly challenging, even with the inclusion of new physics \cite{Miro:2024fid}. As such, it is interesting to consider the additional effect of $B_c$ mesogenesis.

\subsection{\boldmath Charged $B_c$ Mesogenesis}\label{sec:res:charged}
To estimate the contribution arising from $B_c^+ \to B^+ M^0$ decays, we use the branching ratio predictions obtained in Sec.~\ref{sec:BRpred}. We recall that, since the final state $B$ meson is charged, no meson-antimeson oscillations  occur and therefore these modes are also not suppressed by electron rescattering. Consequently, the only source of asymmetry in these channels originates from direct CP violation, defined in Eq.~\eqref{eq:adircharm}. 

As mentioned, the CF and DCS decays $B_c^+ \to B^+ \bar K^{(*)0}$ and $B_c^+ \to B^+ K^{(*)0}$ do not exhibit CP violation in the SM. Therefore, although included above for completeness, these decays do not contribute to the baryon asymmetry.

At the moment, CP violation in SCS $B_c^+\to B$ modes has not yet been observed. To estimate the contribution to the baryon asymmetry, and to assess the baseline viability of the mechanism, we assume that the CP asymmetry in all relevant $B_c^+ \to BM$ channels is of comparable size and shares the same sign. We collectively denote this CP asymmetry by $\mathcal{A}_\mathrm{CP}^\mathrm{dir}(B_c^+ \to B^+ M^0)$. This assumption can be improved once data is available or with advancement in the theoretical calculations.

The contribution from $B_c^+\to B^+$ decays defined in Eq.~\eqref{eq:chargedBc} at $T_R=25\; \mathrm{MeV}$ is then given by 
\begin{align}\label{eq:Yplus}
 Y_\mathfrak{B}^{B_c^+ \to B^+}&=6.5\times 10^{-4}\;\mathcal{B}(B \to \mathfrak{B}\xi \phi)\;\sum_{M^0}\mathcal{B}(B_c^+\to B^+ M^0)\mathcal{A}_\mathrm{CP}^\mathrm{dir}(B_c^+\to B^+ M^0).
\end{align}
We note that the prefactor is linearly dependent on the reheating temperature, therefore higher temperature scenarios are favoured. We further discuss this in Sec.~\ref{sec: combined model}.
Using our results for the branching ratio determinations in Table~\ref{tab:BctoB_BRs} we obtain:
\begin{align}
 Y_\mathfrak{B}^{B_c^+ \to B^+,\mathrm{\rm fac}}&=(3.1 \times 10^{-10}) \times \mathcal{A}_\mathrm{CP}^\mathrm{dir}(B_c^+ \to B^+ M^0)\ , \\
 Y_\mathfrak{B}^{B_c^+ \to B^+,\mathrm{\rm dd}}&=(1.0 \times 10^{-9}) \times \mathcal{A}_\mathrm{CP}^\mathrm{dir}(B_c^+ \to B^+ M^0),
\end{align}
where we use the benchmark $\mathcal{B}(B \to \mathfrak{B}\xi \phi) = 10^{-3}$ and the leading-order and data-driven approach, respectively. The observed difference between the two approaches is driven by the difference in the branching ratio determinations of the $B_c^+ \to B^+ \pi^0$ and $B_c^+ \to B^+ \rho^0$ modes.

Comparing $Y_\mathfrak{B}^{B_c^+\to B^+}$ with neutral $B$ mesogenesis requires an estimate for $\mathcal{A}_\mathrm{CP}^\mathrm{dir}$. The direct CP asymmetry, given in Eq.~\eqref{eq:adircharm}, depends on the relative strong-phase differences between the CKM-leading (tree) amplitudes $\mathcal{T}_i$ and the CKM-suppressed (penguin) amplitudes $\mathcal{P}_i$, which are challenging to calculate.  Assuming ${\rm Im}\; \mathcal{P}_i/\mathcal{T}_i\sim \mathcal{O}(1)$, gives $\mathcal{A}_\mathrm{CP}^\mathrm{dir}\sim 10^{-3}$ as a typical size of the CP asymmetry in charm. At the same time, the penguin amplitude could be enhanced with respect to the tree contribution. Measurements of CP violation in hadronic $D\to M_1M_2$ decays show CP asymmetries at the $10^{-3}-10^{-2}$ level, e.g. in $D^0 \to K^-K^+$ \cite{LHCb:2022lry, LHCb:2019hro} or $D^0\rightarrow K_S^0K_S^0$ \cite{LHCb:2025KS0KS0}, which allow for such an enhancement. A wide variety of theoretical studies have been performed to explain these results, but {\it charm CP violation} remains relatively poorly understood. More data in both the $D$ and $B_c$ systems would further shed light on this.

Allowing for percent level CP asymmetries gives $Y_\mathfrak{B}^{B_c^+ \to B^+}\sim \mathcal{O}(10^{-11})$. Comparing with \eqref{eq:res0} and \eqref{eq:res01} requires information on the relative production fractions of the $B_c$ mesons over the $B_{s(d)}^0$. For sufficiently large $f_c$ production fractions, the $B_c^+ \to B^+$  decays could be comparable in magnitude to that arising from neutral $B$ mesogenesis, potentially canceling negative contributions in the $B_d^0$ sector. We discuss the production fractions further for the combined $B$ mesogenesis model below.

\subsection{\boldmath Hybrid $B_c$ Mesogenesis} \label{sec:res:hybrid}
We continue discussing the new hybrid $B_c$ mesogenesis contributions, where the $B_c$ decays to neutral mesons. The decays involving $B_s^0$ meson mixing are the $B_c^+\to B_s^0 \pi^+$ and $B_c^+\to B_s^0 K^+$ decays and their vector meson counterparts, while those involving $B_d^0$ meson mixing are the $B_c^+\to B_d^0\pi^+$ and $B_c^+\to B_d^0K^+$ decays plus their vector modes. Of the possible channels, only the $B_c^+\to B_s^0 \pi^+$ and $B_c^+\to B_s^0 \rho^+$ decays are CF. These decays are mediated by colour-allowed tree topologies and as such are predicted to have the largest branching ratio, at the level of a few percent. As this mode does not have direct CP violation, the only contribution to $Y_\mathfrak{B}$ comes from mixing effects. Due to their CKM suppression, the SCS modes which exhibit direct CP violation, are predicted to be about an order of magnitude smaller. Finally, the DCS $B_c^+\to B_d^0 K^+$ mode has a small effect as its branching ratio is about $10^{-4}$ and only contributes due to $B_d^0$ mixing. From Eq.~\eqref{eq:DeltaBcneutral}, we note that the decays without a direct CP asymmetry contribute through the time-integrated asymmetry, similar to the neutral decays, but now multiplied with the respective $B_c^+\to B_q^0M^+$ branching ratio.

Adding the predictions in Table~\ref{tab:BctoB_BRs}, we find 
\begin{align}\label{eq:bradd}
    \mathcal{B}^{\rm fac (dd)}({B_c^+ \to B_d^0 M^+}) &= 3.5\; (5.3)\times 10^{-3}\\
    \mathcal{B}^{\rm fac (dd)}({B_c^+ \to B_s^0 M^+}) &= 5.1 \;(4.7)\times 10^{-2} \ ,
\end{align}
in the leading-order factorization (data-driven) approach. Overall, the results from both approaches are in good agreement. As discussed, we do not assign an uncertainty to these estimates, although the difference between the two models may be interpreted as a contribution to the uncertainty\footnote{For the $\eta$ and $\eta'$ modes, we always use the leading-order approach.}.
We note that the $B_s^0$ modes contribute about an order of magnitude more than the $B_d^0$ modes.

We recall from Sec.~\ref{sec: estimates from neutral B mesogenesis} that the neutral mixing contribution from $B_d^0$ modes is negative and suppressed relative to the $B_s^0$ ones due to electron scattering (see Eq.~\eqref{eq:res0} and Eq.~\eqref{eq:res01}). Interestingly, this hierarchy is even more pronounced here due to larger branching ratios of the $B_c^+\to B_s^0$ modes.

The additional term in the hybrid scenario, proportional to the direct CP asymmetry present in SCS modes, is given by the difference of the time-integrated asymmetries. Numerically, we find at $T_R= 25\,\mathrm{MeV}$
\begin{align}
\mathcal{A}_\mathrm{CP}^\mathrm{TI}(B^0_d\to f)-\mathcal{A}_\mathrm{CP}^\mathrm{TI}(\bar B_d^0\to f)&= 2.0\times 10^{-3} \, \label{eq:diffACPTIBd}\\ 
\mathcal{A}_\mathrm{CP}^\mathrm{TI}(B^0_s\to f)-\mathcal{A}_\mathrm{CP}^\mathrm{TI}(\bar B_s^0\to f)&=3.3 \times 10^{-4} \label{eq:diffACPTIBs}
\end{align}
which, as discussed, is effectively independent of $a_q$. We observe that the difference between the $B_d^0$ and $B_s^0$ modes is an order of magnitude. In vacuum, one would expect a much stronger suppression of the $B_s^0$ contribution due to the $1/(1+x_q^2)$ dependence, driven by the rapid oscillation (large $x_s$). However, the inclusion of the scattering effects significantly alters the time-integrated asymmetries, effectively smoothing the otherwise strong vacuum suppression in the $B_s^0$ sector.

We can now combine the contributions proportional to the direct CP asymmetry with those arising from mixing. As discussed, we do not have any measurement or prediction of CP violation in the $B_c$ system. Therefore, we assume all $\mathcal{A}_\mathrm{CP}^\mathrm{dir}$ in $B_c^+\to BM$ decays to be universal in magnitude and sign. Combining both contributions, we find
\begin{equation}
\begin{aligned}
Y_\mathfrak{B}^{B_c^+ \to B_d^0} = 1.3\times 10^{-3}\mathcal{B}(B_d^0\to \mathfrak{B}\xi\phi)&\left\{1.0\times 10^{-3}a_d\sum_{M^+}\mathcal{B}(B_c^+ \to B_d^0M^+) \right.\\
    &\left.+ \sum_{M^+}\mathcal{B}(B_c^+ \to B_d^0M^+)\mathcal{A}_\mathrm{CP}^\mathrm{dir}(B_c^+\to B_d^0 M^+)\right\}.
    \end{aligned}
    \label{eq: YB BcBd general}
\end{equation}
\begin{equation}
\begin{aligned}
Y_\mathfrak{B}^{B_c^+ \to B_s^0} = 7.3 \times 10^{-4}\mathcal{B}(B_s^0\to \mathfrak{B}\xi\phi)&\left\{0.4\times a_s\sum_{M^+}\mathcal{B}(B_c^+ \to B_s^0M^+) \right.\\
    &\left.+ \sum_{M^+}\mathcal{B}(B_c^+ \to B_s^0M^+)\mathcal{A}_\mathrm{CP}^\mathrm{dir}(B_c^+\to B_s^0 M^+)\right\},
\end{aligned}
\label{eq: YB BcBs general}
\end{equation}
where the uncertainties from $x_q$ and $y_q$ are again negligible. 

Using the branching ratio values from Table~\ref{tab:BctoB_BRs} and $a_q^{\rm theo}$ in Eq.~\eqref{eq:atheo} as input, we obtain from Eqs.~\eqref{eq: YB BcBd general} and~\eqref{eq: YB BcBs general}:
\begin{align}
    Y_{\mathfrak{B}}^{B_c^+\to B_d^0,\mathrm{fac,theo}}&= (-2.4 \pm 0.2)\times 10^{-15} +  4.4\times 10^{-9}\mathcal{A}_\mathrm{CP}^\mathrm{dir}(B_c^+ \to B_d^0M^+) \ , \\
    Y_\mathfrak{B}^{B_c^+\to B_s^0,\mathrm{fac,theo}}&=  (3.3\pm 0.2)\times 10^{-13} + 1.7 \times 10^{-9}\mathcal{A}_\mathrm{CP}^\mathrm{dir}(B_c^+ \to B_s^0M^+) \ ,
\end{align}
for the $B_d^0$ and $B_s^0$, respectively using the leading-order factorization approach. We find similar results for the data-driven approach. We observe a strong hierarchy between the mixing-induced and direct CP violating terms, especially in the $B_d^0$ system, which is caused by a strong numerical cancellation in $\mathcal{A}_\mathrm{CP}^\mathrm{TI}(\langle B_d^0 \to f \rangle)$, which significantly suppresses the mixing-induced term.

Assuming a direct asymmetry of $\mathcal{O}(10^{-3}-10^{-2})$, the contribution from the direct CP asymmetry term is dominant compared to the mixing-induced component, in both the $B_d^0$ and $B_s^0$ sectors.  Therefore, these results provide a compelling additional motivation for measurements of direct CP violation in $B_c$ decays, for which both the $B_c^+\to B_s^0 K^+$ and $B_c^+\to B_d^0 \pi^+$ are prime candidates.

 It is worth noting that $Y_{\mathfrak{B}}^{B_q^0}$ and $Y_\mathfrak{B}^{B_c^+\to B_q^0}$ can be of comparable orders of magnitude. Ultimately, their relative contribution is proportional to their respective fragmentation fractions. 

For completeness, we also quote the results using $a_q^{\rm exp}$ from Eq.~\eqref{eq:aexp} in Eqs.~\eqref{eq: YB BcBd general} and~\eqref{eq: YB BcBs general}:
\begin{align}
    Y_\mathfrak{B}^{B_c^+\to B_d^0,\mathrm{fac,exp}}&= (-9.6 \pm 7.7)\times 10^{-15} + 4.4 \times 10^{-9}\mathcal{A}_\mathrm{CP}^\mathrm{dir}(B_c^+ \to B_d^0M^+) \ , \\
 Y_\mathfrak{B}^{B_c^+\to B_s^0,\mathrm{fac,exp}}&=(-0.9 \pm4.1)\times 10^{-11}  + 1.7 \times 10^{-9}\mathcal{A}_\mathrm{CP}^\mathrm{dir}(B_c^+ \to B_s^0M^+) \ ,
\end{align}
in the leading-order factorization approach and similar results in the data-driven approach.

To conclude, the contribution from direct CP violation is crucial for both the charged and hybrid mechanisms, and can significantly alter the total baryon yield. We discuss the dependence of these results on the reheating temperature and the detailed interplay between neutral and $B_c$ mesogenesis in the following section.

% ===========================================

\section{Unified \boldmath $B$ mesogenesis} \label{sec: combined model}

\begin{figure}[t]
    \centering
    \includegraphics[width=1\textwidth]{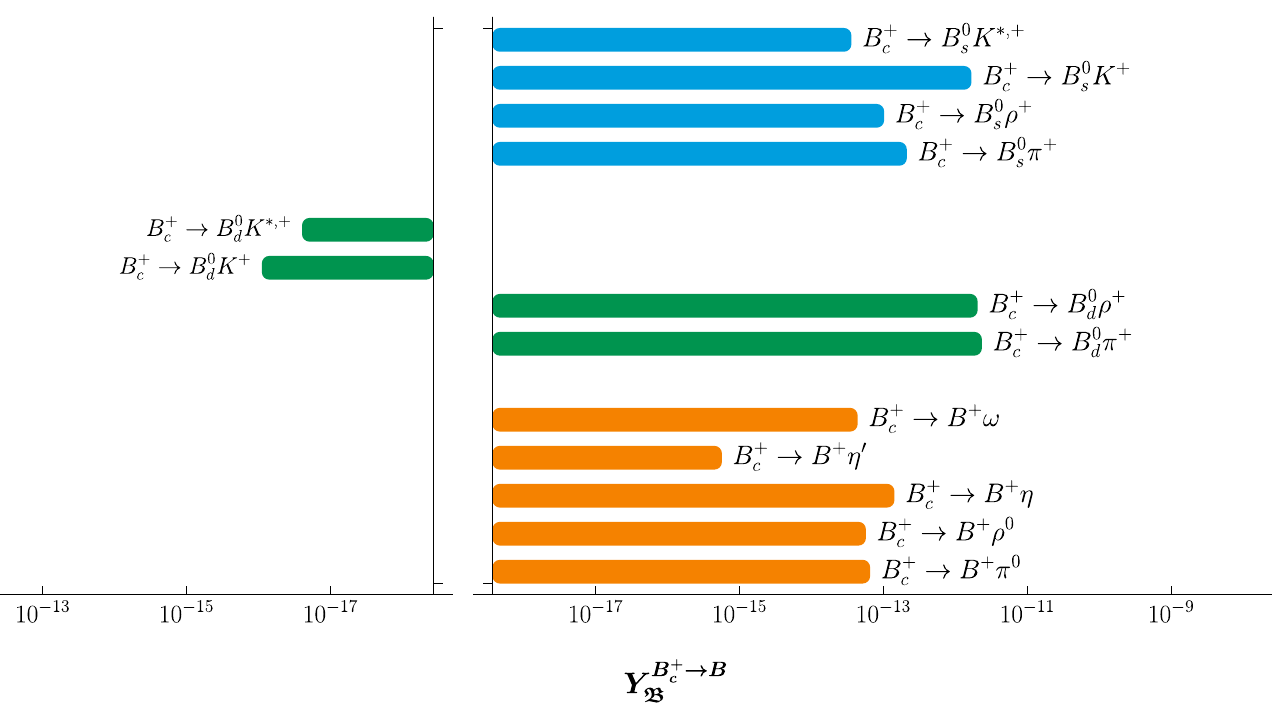}
    \caption{Estimates of the individual contributions to the baryon asymmetry $Y_{\mathfrak{B}}^{B_c^+\to B}$ at $T_R=25 \,\mathrm{MeV}$ from each $B_c^+ \to BM$ decay channel using $a_q^{\rm theo}$ in Eq.~\eqref{eq:atheo}. We fix $\mathcal{A}_\mathrm{CP}^\mathrm{dir}(B_c^+ \to BM) = 10^{-3}$, $\mathcal{B}(B \to \mathfrak{B}\xi\phi)=10^{-3}$ and use our leading-order approach for the calculation of the $B_c^+\to BM$ branching ratios. Orange bars correspond to $B_c^+ \to B^+ M^0$ modes, green bars to $B_c^+ \to B_d^0 M^+$ and blue bars to $B_c^+ \to B_s^0M^+$. In order to obtain the total baryon asymmetry these numbers must be multiplied by the fragmentation fraction $f_c$. }
    \label{fig:Bc indv contributions}
\end{figure}

We first discuss the total impact of $B_c$ mesogenesis,
\begin{equation}
    Y_\mathfrak{B}^{B_c^+\to B} \equiv Y_\mathfrak{B}^{B_c^+\to B^+} + Y_\mathfrak{B}^{B_c^+\to B_q^0},
\end{equation}
to the baryon asymmetry using the results from Sec.~\ref{sec:res:charged} and Sec.~\ref{sec:res:hybrid}. 
In Fig.~\ref{fig:Bc indv contributions}, we show the contribution of each of the $B_c^+ \to B M$ modes using the estimated branching ratios in the leading-order approach. Note that each mode must be weighted by the fragmentation fraction $f_c$ when computing the total baryon asymmetry. The contributions from the $B_c^+\to B^+ M^0$ channels, which are proportional to the direct CP violation, are depicted in orange.  The contributions of the $B_c^+ \to B_d^0 M^+$ and $B_c^+ \to B_s^0 M^+$ channels are given in green and blue, respectively, for which we use the $a_q^{\rm theo}$ given in Eq.~\eqref{eq:atheo}. In addition, we assume $\mathcal{A}_\mathrm{CP}^\mathrm{dir}(B_c^+ \to BM) = 10^{-3}$ for all modes. 
From Fig.~\ref{fig:Bc indv contributions}, we observe:
\begin{itemize}
    \item {$\boldsymbol{ B_c^+ \to B^+ M^0}$}: almost all channels contribute comparably, except for the $\eta'$ mode due to its suppressed branching ratio. The largest contribution comes from the $B_c^+ \to B^+ \eta$ decay. 
    \item $\boldsymbol{B_c^+ \to B_s^0 M^+}$: in this class of modes all channels contribute at the same level. Specifically, the lower branching ratios of the final states containing a $K$ or $K^*$ are compensated by the direct CP violation contribution. Measurements of CP violation in these channels are of key importance to further study $B_c$ mesogenesis.
    \item $\boldsymbol{B_c^+ \to B_d^0 M^+}$: this is the only class with modes that have negative contributions to the asymmetry, which arise from the minus sign of $a_d$. The DCS modes $K^+$ and $K^{*,+}$ only contribute through mixing and are therefore negative. In contrast, the SCS modes $\pi^+$ and $\rho^+$ receive sizeable contributions from $\mathcal{A}_\mathrm{CP}^\mathrm{dir}$, which compensates for the negative mixing effects. Assuming $\mathcal{A}_\mathrm{CP}^\mathrm{dir} = 10^{-3}$, these decays give the largest contribution to $Y_\mathfrak{B}^{B_c^+ \to B}$. Experimental measurements of CP violation in the $B_c^+\to B_d^0 \pi^+(\rho)$ are therefore strongly motivated. 
\end{itemize}

\noindent
In order to find the overall effect of $B$ mesogenesis, we combine all contributions as in Eq.~\eqref{eq: evolution individual baryon asymmetries}.

The relative impact of the $B_c$ versus neutral mesogenesis contributions depends on the fragmentation fractions $f_c$ and $f_q$, respectively. Estimations and experimental measurements exist in the context of collider environments,
\begin{equation}\label{eq:fs}
    f_d^\mathrm{coll.} = f_u^\mathrm{coll.} = 0.407~\text{\cite{HeavyFlavorAveragingGroupHFLAV:2024ctg}}, \hspace{1cm} f_s^\mathrm{coll.} = 0.101~\text{\cite{HeavyFlavorAveragingGroupHFLAV:2024ctg}}, \hspace{1cm} f_c^\mathrm{coll.} = 2.6 \times 10^{-3}~\text{\cite{LHCb:2019tea}}.
\end{equation}
Using these collider values, the total baryon asymmetry at $T_R=25 \,\mathrm{MeV}$ is:
\begin{align}
    Y_\mathfrak{B}^\mathrm{fac,exp}&= (-1.8 \pm 8.1 )\times 10^{–11} + 1.7 \times 10^{-11}\mathcal{A}_\mathrm{CP}^\mathrm{dir}(B_c^+\to BM) \ ,\\
    Y_\mathfrak{B}^\mathrm{fac,theo}&= (3.8 \pm 0.4 )\times 10^{–13} + 1.7 \times 10^{-11}\mathcal{A}_\mathrm{CP}^\mathrm{dir}(B_c^+\to BM) \ ,
\end{align}
where we used $a_{q}^{\rm theo}, \, a_q^\mathrm{exp}$ and the branching ratios obtained in the factorization approach. The tiny fragmentation function $f_c$ reduces the effect of the direct CP violation term of the $B_c$ sector. We observe that for $a_q^{\rm theo}$ it is challenging to produce sufficient baryon asymmetry from the mixing alone, as already discussed in the neutral case. With our estimates, getting $Y_\mathfrak{B} \sim 10^{-11}$, requires charm CP violation at $\mathcal{O}(1)$. For the experimental case similarly large values for direct CP violation are necessary to compete with the contribution from neutral $B$-meson mixing. Obtaining more than percent level CP violation would require a strong enhancement of the penguin amplitude with respect to the tree-level amplitudes, which seems out of line with SM expectation and would probably require new physics effects. While in that case $B$ mesogenesis would not explain the baryon asymmetry of the universe through SM CP violation alone, it still remains an interesting mechanism to convert CP violation in the $B$ sector into a baryon asymmetry.
\begin{figure}[t]
\centering
\begin{subfigure}[b]{0.49\textwidth}
\includegraphics[width=1\textwidth]{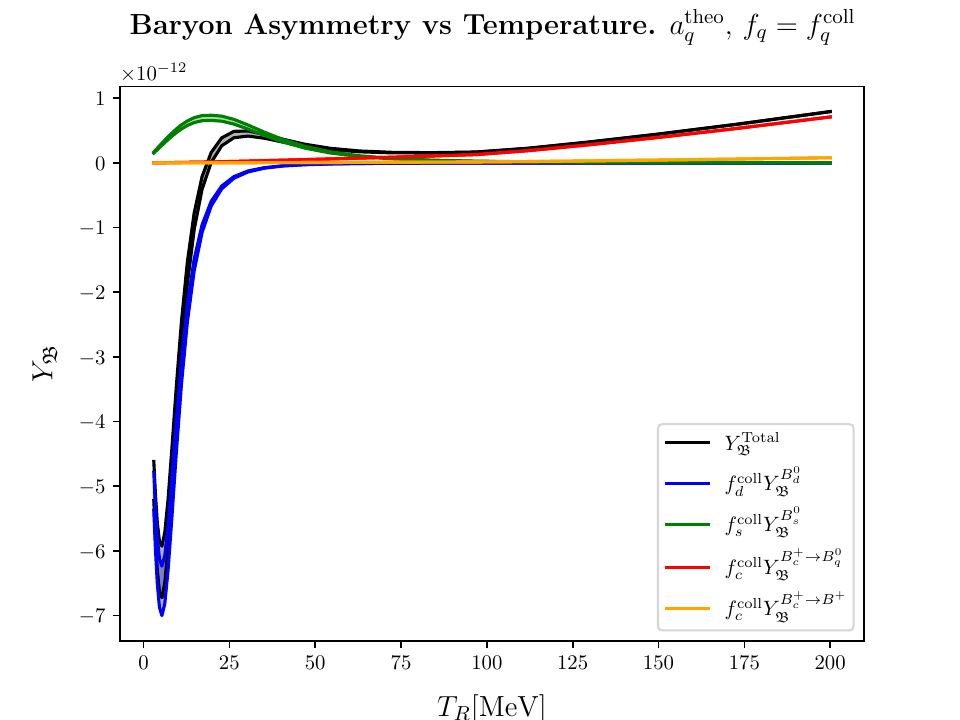}
  \caption{}
\end{subfigure}
\begin{subfigure}[b]{0.49\textwidth}
\includegraphics[width=1\textwidth]{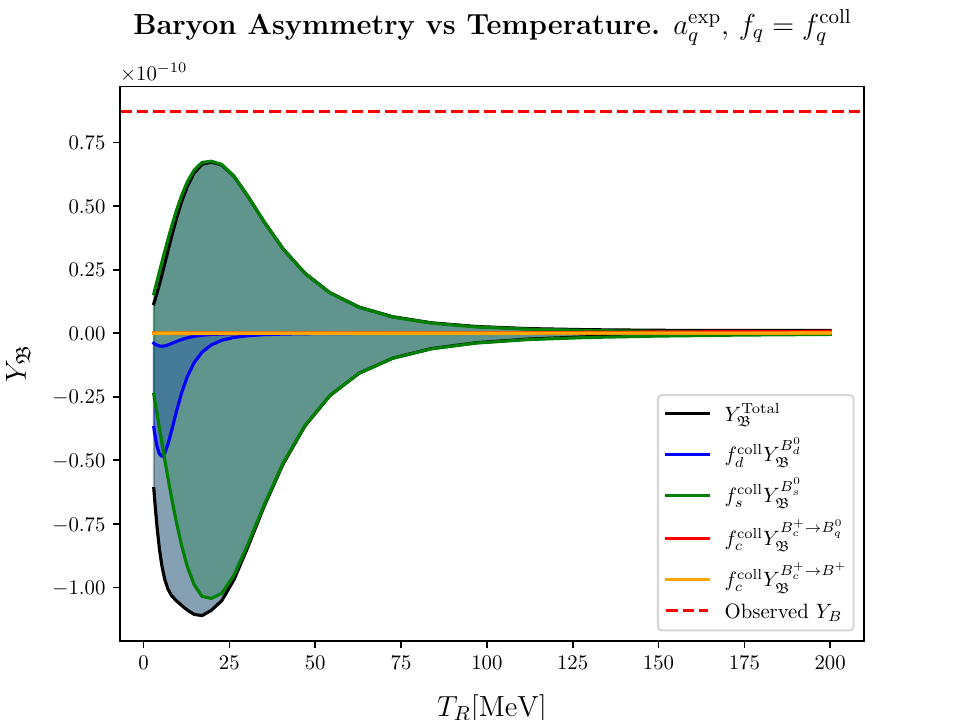}
\caption{}
\end{subfigure}
\caption{Total baryon asymmetry $Y_\mathfrak{B}$ as a function of the reheating temperature $T_R$, assuming collider fragmentation fractions, using theoretical (left) and experimental (right) values for $a_q$. The black line represents the total baryon asymmetry, while the coloured lines represent the individual contributions: the neutral $B_d^0$ (blue) and $B_s^0$ (green) mixing mechanisms, and $B_c$ decays into neutral $B_q^0$ mesons (red) and into charged $B^+$ mesons (yellow). The horizontal dashed line in Figure b) indicates the observed baryons asymmetry value.}
\label{fig:YB_TR}

\end{figure}

Above, we presented our numerical results at fixed reheating temperature $T_R=25 \,\mathrm{MeV}$. In Fig.~\ref{fig:YB_TR}, we show the contribution of each mechanism to the total baryon asymmetry as a function of $T_R$ using the production fractions in Eq.~\eqref{eq:fs}. While the mixing-induced component dominates for low reheating temperatures $(T_R\lesssim 75 \,\mathrm{MeV}$), the terms originating from the direct CP asymmetry become the dominant contribution at higher temperatures. In particular, the direct CP asymmetry contributions from the hybrid mechanism dominate with respect to the charged ones. These contributions are linearly dependent on $T_R$ since they are not suppressed by electron rescattering. However, even at high $T_R$, these contributions remain insufficient to reproduce the observed baryon asymmetry when adopting the $B_c$ production fraction as measured at colliders and the quoted benchmark value for $\mathcal{A}_\mathrm{CP}^\mathrm{dir}(B_c^+ \to BM)$. Specifically, at high $T_R$, for our estimate of branching ratios, we would require $\mathcal{A}_\mathrm{CP}^\mathrm{dir}(B_c^+ \to BM)$ around $(10-15)\%$ and thus a penguin over tree ratio of around 10 to 15 to explain the baryon asymmetry.  We recall that $T_R<\Lambda_\mathrm{QCD}$ to ensure hadronization before the quarks decay.

However, early-universe conditions may enhance the production of heavier mesons, leading to larger $f_c$ than observed at colliders. This enhancement could be driven by new physics mechanisms, such as additional scalar couplings; however, we do not explore these specific scenarios here. Nonetheless, we consider it interesting to allow $f_c$ to vary and to determine the values needed to match the observed baryon asymmetry. We consider two separate scenarios for the light-quark fragmentation fractions:  $f_u=f_d=f_s$ and a hierarchy similar to collider measurements by setting $f_u=f_d=4f_s$. Additionally, we neglect any $b \to$ baryon hadronization processes by imposing $f_u+f_d+f_s+f_c=1$ in both scenarios.

\begin{figure}[t]
\centering
\begin{subfigure}[b]{0.49\textwidth}
\includegraphics[width=1\textwidth]{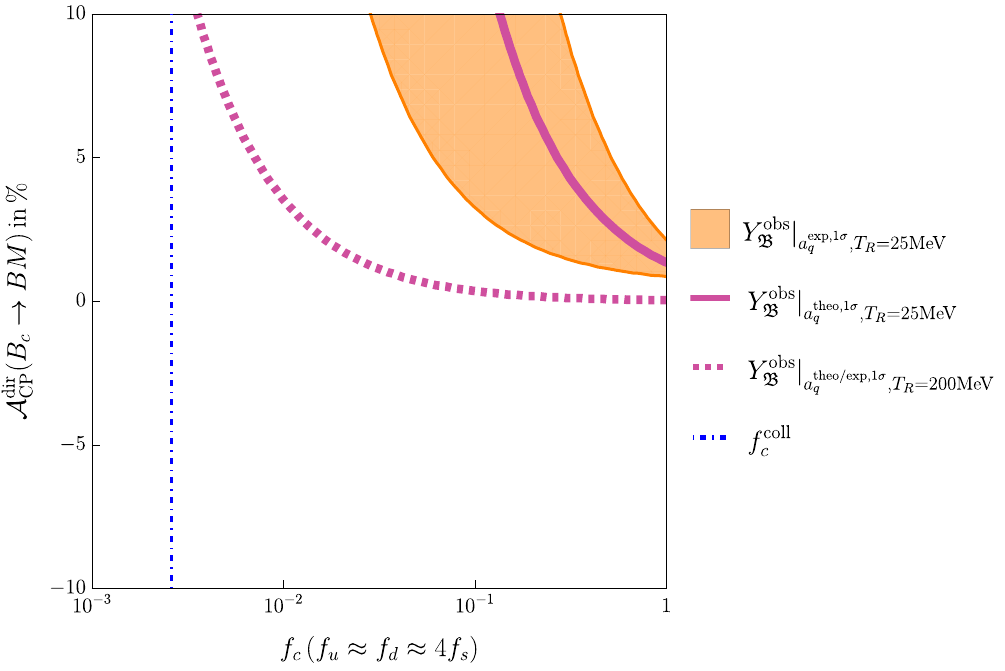}
  \caption{}
\end{subfigure}
\begin{subfigure}[b]{0.49\textwidth}
\includegraphics[width=1\textwidth]{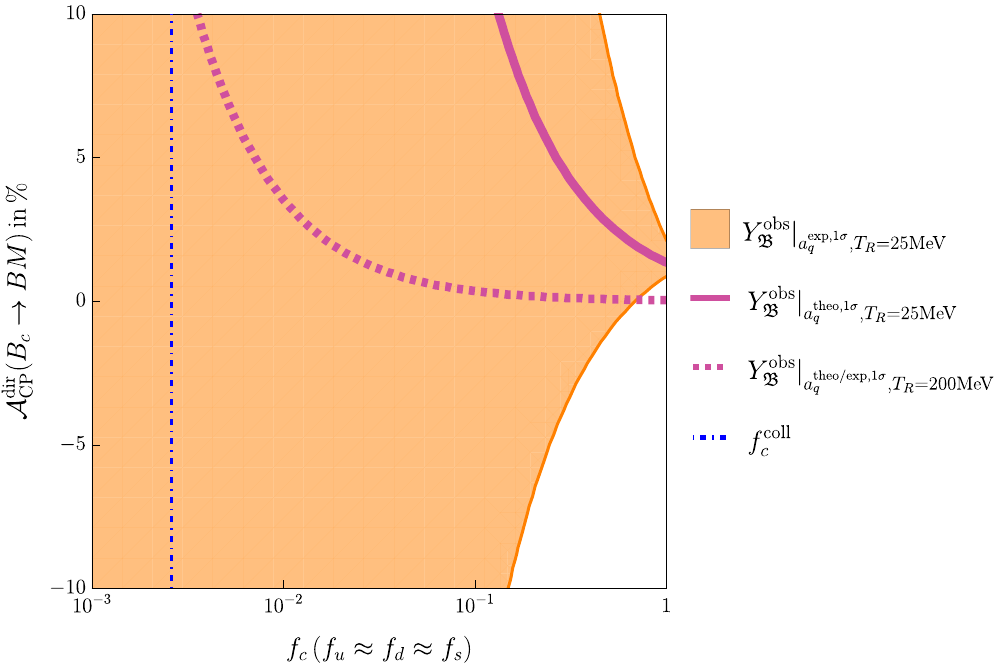}
\caption{}
\end{subfigure}
    \caption{Viability of the unified $B$ mesogenesis framework as a function of the fragmentation fraction $f_c$ and $\mathcal{A}_\mathrm{CP}^\mathrm{dir}$ for $T_R=25\,\mathrm{MeV}$ (solid) and $T_R=200 \,\mathrm{MeV}$ (dashed). We set $\mathcal{B}(B\to f)=10^{-3}$ and use leading-order factorization to estimate the $B_c$ branching ratios. Orange (pink) solid regions correspond to experimental (theoretical) determinations of the semileptonic asymmetries $a_q$ at $1\sigma$, using a) collider-like hierarchy for the light-quark fragmentation fractions, and b) symmetric fractions ($f_u=f_d=f_s$). Note that at $T_R=200\,\mathrm{MeV}$, the experimental and theoretical regions coincide because the direct CP contribution dominates which is independent of $a_q$.}
    \label{fig:combined results}
\end{figure}

The viability of unified $B$ mesogenesis as a function of the fragmentation fraction $f_c$ and the direct CP asymmetry $\mathcal{A}_{\rm CP}^{\rm dir}$ is shown in Fig.~\ref{fig:combined results}. The displayed contours represent the parameter space where the predicted asymmetry exactly matches the $Y_\mathfrak{B}^\mathrm{obs}$ given in Eq.~\ref{eq: baryon asymmetry}. Regions towards the top-right, i.e. larger values of $f_c$ and $\mathcal{A}_\mathrm{CP}^\mathrm{dir}$, would result in a total baryon asymmetry that overshoots the observed value. The solid regions shown in orange (pink) correspond to the experimental (theoretical) determination of the semileptonic asymmetry at $1\sigma$, $a_q^{\mathrm{exp},1\sigma} (a_q^{\mathrm{theo},1\sigma})$ at $T_R=25 \, \mathrm{MeV}$. The dashed contour indicates viable $B$ mesogenesis regions for $T_R=200 \,\mathrm{MeV}$. In this high-$T_R$ scenario, predictions adopting $a_q^\mathrm{exp}$ and $a_q^\mathrm{theo}$ overlap since the asymmetry is entirely driven by the $B_c$ contribution, making the final result independent of the semileptonic asymmetry.

Shifting from the collider-like hierarchy ($f_d \approx 4f_s$) to the symmetric case ($f_u=f_d=f_s$) in which the production of $B_s^0$ mesons is enhanced, strongly boosts the mixing contribution at low reheating temperatures. However, this sensitivity to the fragmentation fractions only holds when the neutral channel provides a sizeable effect, namely when adopting the $1\sigma$ experimental values for the semileptonic asymmetries. If theoretical predictions are adopted instead, the neutral contribution becomes negligible and the total baryon asymmetry is dominated by the contribution from $B_c$ mesons. Consequently, the fragmentation fractions of $B_d^0$ and $B_s^0$ mesons have a negligible impact on the final asymmetry in this scenario.

As said, the direct CP asymmetry part of $B_c$ mesogenesis plays a key role for higher reheating temperatures. Specifically, at $T_R = 200$ MeV, we find
\begin{equation}
\begin{split}
    Y_\mathfrak{B}^\mathrm{fac}= &\left\{1.1\times 10^{-9} a_d \left[f_d + f_c\sum_{M^+}\mathcal{B}(B_c^+\to B^0_dM^+)\right] + \right.\\
    &\quad1.4\times 10^{-6}a_s\left[f_s + f_c\sum_{M^+}\mathcal{B}(B_c^+\to B^0_sM^+)\right]+ \\
    &\quad 0.02 \,f_c  \sum_{M^0}\mathcal{B}(B_c^+\to B^+M^0)\mathcal{A}_\mathrm{CP}^\mathrm{dir}(B_c^+ \to B^+ M^0)+\\
    & \quad \left.0.04\,f_c\sum_{q,M^+}\mathcal{B}(B_c^+\to B^0_qM^+)\mathcal{A}_\mathrm{CP}^\mathrm{dir}(B_c^+ \to B^0_q M^+)\right\} \mathcal{B}(B\to \mathfrak{B}\xi\phi) 
    \end{split}
\end{equation}
In this high-temperature scenario, $B_c$ mesogenesis is viable if 
\begin{align}\label{eq:BcathighT}
\, &\sum_{M^0}\mathcal{B}(B_c^+\to B^+M^0)\mathcal{A}_\mathrm{CP}^\mathrm{dir}(B_c^+ \to B^+ M^0)+ \nonumber \\
{}& 2\sum_{q,M^+}\mathcal{B}(B_c^+\to B^0_qM^+)\mathcal{A}_\mathrm{CP}^\mathrm{dir}(B_c^+ \to B^0_q M^+) \sim 2\cdot 10^{-3} \left(\frac{2.6\times 10^{-3}}{f_c}\right) , 
\end{align}
using $\mathcal{B}(B\to \mathfrak{B}\xi\phi)=10^{-3}$. From Table~\ref{tab:BctoB_BRs}, we note that the expected branching ratios for the $B_c^+\to B^+M^0$ SCS decay are suppressed. This leaves the $B_c\to B_s^0 K^+$ and $B_c\to B_d^0\pi^+ (\rho^+)$ as the driving modes to generate $B_c$ mesogenesis assuming SM sources of CP violation. Assuming branching ratios at the percent level, we conclude that assuming $f_c$ as at colliders, requires CP asymmetries at the $10\%$ level. As the relation is linear, allowing for a factor $10$ increase in $f_c$, setting it at the percent level, gives viable $B_c$ mesogenesis with CP asymmetries at the percent level. We stress that CP asymmetries at the percent level are compatible with expected ranges for charm CP violation, supporting the viability of the combined scenario for a slightly enhanced $f_c$, mainly for higher reheating temperatures. 

In general, we assumed that the direct CP violation is of the same order and sign in the different modes, which is a strong assumption. Once measurements of the CP asymmetries and branching ratios of the dominant $B_c\to B_s^0 K^+$ and $B_c\to B_d^0\pi^+ (\rho^+)$ modes are available, Eq.~\eqref{eq:BcathighT} can be used to check the viability of $B_c$ mesogenesis at high temperatures. First data on CP violation in $B_c$ decays could potentially show further surprises, similar as in the $D$ sector, where CP asymmetries at the percent level are still allowed by data. Either way, new measurements will certainly provide interesting additional information. Also updated measurements of $a_q^{\rm exp}$ could sharpen the picture even further in the low $T_R$ regime. 

Finally, we stress that $Y_\mathfrak{B}$ is linearly proportional to the baryonic branching ratio of $B$ mesons, for which we used $\mathcal{B}(B\to \mathfrak{B}\xi\phi) \sim 10^{-3}$ for the inclusive branching ratio in line with current experimental limits as discussed above. For more stringent limits, the viable regions of $B$ mesogenesis will shrink accordingly.

\section{Conclusions} \label{sec: conclusions}

$B$ mesogenesis is an interesting framework for explaining the baryon asymmetry of the universe. It provides a concrete mechanism for solving one of the key problems of the SM: translating CP violation into a net baryon asymmetry. While the framework requires additional hypothetical fields, it relies originally on direct and mixing-induced CP violation generated within the SM. In this work, we combined all flavours of $B$ mesogenesis for the first time: we have examined three channels that contribute to the baryon asymmetry: neutral $B$ mesogenesis (via $B_q^0-\bar B_q^0$ oscillations), charged $B$ mesogenesis from $B_c^+ \to B^+ M^0$, and a new hybrid mechanism involving $B_c^+ \to B_q^0 M^+$, followed by neutral meson oscillations.

Regarding neutral mesogenesis, we reviewed the mechanism and provided the time-dependent decay rates including electron scattering. These rates introduce a strong dependence on the reheating temperature, strongly suppressing the neutral mechanism for temperatures above $\sim75\,\mathrm{MeV}$. The viability of neutral $B$ mesogenesis also depends on the relevant fragmentation fractions: obtaining sufficient baryon asymmetry through neutral mesons alone would require an enhanced $B_s^0$ production, or a positive semileptonic asymmetry for $B_d^0$ mesons, which is still allowed from measurements but disfavoured by theoretical predictions. Interestingly, the expected negative $B_d^0$ contribution can be compensated not only by the $B_s^0$ channel, but also by the additional contributions from $B_c$ decays of the unified mechanism.

We considered two additional sources to the asymmetry arising from $B_c$ decays: the previously studied $B_c^+\to B^+ M^0$ channel and the new $B_c^+ \to B_q^0M^+$ mechanism introduced in this work. The impact of these $B_c$ modes depends on the currently unknown branching ratios of the $\mathcal{B}(B_c^+ \to BM)$ decays. We estimated these branching ratios within a factorization framework using two complementary approaches: leading-order factorization and a data-driven approach using $D\to M_1M_2$ decays. We find good consistency between both methods, within an order of magnitude. Measurements of these branching ratios are highly anticipated to confront our approach with data and to improve our theoretical understanding of hadronic decays.

Using our branching ratio estimates, we assess the impact of the two $B_c$ mesogenesis mechanisms. The contributions can be grouped according to the dominant source of CP violation. Decays to neutral $B_s^0$ or $B_d^0$ mesons receive contributions from both mixing-induced and direct CP violation. The direct CP violation dominates the baryon asymmetry for values at the subpercent level. To test if such relatively large values of CP violation are indeed generated within the SM (or beyond), we advocate experimental searches for CP violation specifically in $B_c^+\to B_s^0K^+$, $B_c^+\to B_d^0\pi^+$ and $B_c^+\to B_d^0\rho^+$ decays. These decays are predicted to have the largest branching ratios. As such, they would directly probe the most relevant sources of baryon asymmetry within the $B_c$ sector. We stress that, in general, studies of CP violation in $B_c$ decays are of great interest as they probe charm CP violation in decays with estimated branching ratios at the percent level. These decays would provide interesting complementary tests of SM CP violation compared to $D$ decays. In this work, we adopt a simplified treatment of the direct CP asymmetries for the purpose of our theoretical estimates. A more detailed analysis of both the sign and magnitude of the CP asymmetries in individual channels is left for future work.

Finally, we explored the viability of the unified model across different scenarios for the production fractions. At low reheating temperatures, where the mixing contribution dominates, explaining the observed baryon asymmetry requires $a_s$ to be larger than its current theoretical value and possibly an enhancement in the $B_s$ meson production fraction. In contrast, at higher reheating temperatures, the mixing contribution becomes negligible. In this high-temperature regime, the observed baryogenesis can be successfully achieved with both $f_c$ and $\mathcal{A}_\mathrm{CP}^\mathrm{dir}$ at the percent
level. For $f_c\sim 10^{-3}$, we find that $\mathcal{A}_\mathrm{CP}^\mathrm{dir} \sim (10-15)\%$ would explain the baryon asymmetry. This would require the corresponding penguin over tree ratio to be $10-15$. Whether this is realized in nature could be verified with first experimental studies of CP violation in $B_c^+\to BM$ decays. Enhancements of the branching ratios with respect to our estimates could further lower this estimate by a few percent.

In summary, we have presented first estimates for unified $B$ mesogenesis. Future improvements in the measurement of semileptonic asymmetries and first measurements of the direct CP asymmetries in $B_c$, together with dedicated searches for the hypothetical dark-sector particles and rare $B$ decays to baryons, will provide crucial experimental tests of this framework. In parallel, ongoing theoretical and experimental efforts to better understand fragmentation fractions in the early universe and $B_c$ decay dynamics will be essential to clarify the role of CP violation in these channels and fully exploit their potential in generating baryon asymmetry.

\acknowledgments
We thank M. Escudero for useful discussions. This publication is part of the project Solving Beautiful Puzzles with file number VI.Vidi.223.083 of the research programme Vidi which is financed by the Dutch Research Council (NWO).

\appendix

\newpage

\section{Numerical Solution of Boltzmann Equations}\label{ap: Boltzman eqs}

In this appendix we detail the cosmological framework and the numerical procedure used to solve the system of Boltzmann equations that govern the evolution of the energy densities and the generation of the baryon asymmetry. We consider a scenario where the early universe is initially dominated by a heavy particle $\Phi$ with mass $M_\Phi=100 \,\mathrm{GeV}$.
The coupled system of differential equations describing the evolution of the energy density of the parent particle $\rho_\Phi$, the radiation energy density $\rho_R$, and the number densities $n_X$ of the different baryon-asymmetry contributions from $B$ meson sectors is given by:
\begin{align}
    \frac{d}{dt}\rho_\Phi + 3H\rho_\Phi &= -\Gamma_\Phi \rho_\Phi \,, \\
    \frac{d}{dt}\rho_R + 4H\rho_R &= +\Gamma_\Phi \rho_\Phi \,, \\
    \frac{d}{dt}n_X + 3Hn_X &= \frac{1}{2} \Delta_X \Gamma_\Phi^b n_\Phi \,,
\end{align}
where $n_\Phi = \rho_\Phi / M_\Phi$. The index $X \in \{B_q^0, B_c \to B^+, B_c^+ \to B_q^0\}$ labels the different channels contributing to the total baryon asymmetry and $\Delta_X$ represents the effective baryon asymmetry factor per decay in that specific channel, which encapsulates both direct and mixing-induced CP-violating contributions. The expansion of the universe is governed by the dynamic Hubble parameter $H$, defined as:
\begin{equation}
    H = \frac{1}{\sqrt{3} M_{\mathrm{Pl}}} \sqrt{\rho_\Phi + \rho_R} \,,
\end{equation}
where $M_{\mathrm{Pl}} \approx 2.436 \times 10^{18}~\mathrm{GeV}$ is the reduced Planck mass. The decay width $\Gamma_\Phi$ is determined by fixing the reheating temperature $T_R$ through the relation:
\begin{equation}
    \Gamma_\Phi = 3 H(T_R) = 3 \sqrt{\frac{2\pi^2}{90} g_*(T_R) \frac{T_R^4}{M_{\mathrm{Pl}}^2}} \,,
\end{equation}
where $g_*(T)$ is the effective number of relativistic degrees of freedom at temperature $T$, which we take from \cite{Laine:2015kra}.
To solve the system numerically and eliminate the explicit dependence on the universe's expansion terms ($3H, 4H$), we define the variable $x \equiv \ln(a/a_i)$, where $a_i$ is the initial scale factor. We introduce the dimensionless comoving variables:
\begin{equation}
    C_\Phi(x) \equiv \frac{\rho_\Phi(x) a^3}{\rho_{\Phi,i} a_i^3} \,, \quad 
    C_R(x) \equiv \frac{\rho_R(x) a^4}{\rho_{\Phi,i} a_i^4} \,, \quad 
    C_X(x) \equiv \frac{n_X(x) a^3}{n_{\Phi,i} a_i^3} \,,
\end{equation}
where $\rho_{\Phi,i}$ and $n_{\Phi,i} = \rho_{\Phi,i}/M_\Phi$ are the initial energy and number densities of the $\Phi$ field, respectively. 
Using the relation $d/dt = H \, d/dx$, the physical Boltzmann equations map onto a compact system of first-order ordinary differential equations:
\begin{align}
    \label{dC_Phi} \frac{dC_\Phi}{dx} &= -\frac{\Gamma_\Phi}{H(x)} C_\Phi \,, \\
    \label{dC_R} \frac{dC_R}{dx} &= \frac{\Gamma_\Phi}{H(x)} C_\Phi e^x \,, \\
    \label{dC_X} \frac{dC_X}{dx} &= \frac{1}{2} \Delta_X \frac{\Gamma_\Phi}{H(x)} C_\Phi \,,
\end{align}
where the physical energy densities needed to evaluate the Hubble rate at any integration point are reconstructed via:
\begin{equation}
    \rho_\Phi(x) = \rho_{\Phi,i} C_\Phi(x) e^{-3x} \,, \quad \text{and} \quad \rho_R(x) = \rho_{\Phi,i} C_R(x) e^{-4x} \,.
\end{equation}
The instantaneous temperature of the radiation bath $T(x)$ is computed from the physical radiation density as $T = [30 \rho_R / (\pi^2 g_*(T))]^{1/4}$. This temperature tracking is crucial because the effective asymmetry factors $\Delta_X$ for the neutral mesons depend on the electron rescattering rate, which scales strongly with temperature according to Eq.~\eqref{eq: scattering electrons}.
This rate modifies the standard neutral meson mixing parameters into effective ones, altering the time-integrated asymmetries dynamically throughout the cosmic evolution, as shown in Section~\ref{sec:intro_neutmeso}.
The system of equations is integrated from $x=0$ to $x=30$, when the $\Phi$ field has completely decayed. Assuming that the early universe is initially fully dominated by the $\Phi$ field, we impose the following initial boundary conditions
\begin{equation}
    C_\Phi(0) = 1.0 \,, \quad C_R(0) = 10^{-8} \,, \quad C_X(0) = 0.0 \,.
\end{equation}
At the end of the integration, the final comoving values $C_X(x_f)$ are used to reconstruct the physical yields via $Y_X = C_X(x_f) e^{-3x_f} n_{\Phi,i} / s_f$, where $s_f$ is the final entropy density. The total baryon asymmetry of the universe is then simply obtained by summing over all active channels weighted by the corresponding fragmentation fractions, as given by Eq.~\eqref{eq: individual Y's}.

\section{CKM subleading terms}\label{ap:CKM_sublead}
The subleading topological contributions to the amplitudes, $\mathcal{P}_i$, are listed in Table~\ref{tab:BctoB_subleadtopologies}. Here we defined as before $P_{bD} = P_b-P_D$ for $D=s,b$, and equivalently for the other topologies. The colour-allowed EW penguin and QCD penguin diagrams are illustrated in Fig.~\ref{fig: penguin diagrams}. The $\mathcal{P}_i$ amplitudes also include an annihilation topology $A$, absent in the leading contributions. Note that these processes do not involve exchange or penguin-annihilation topologies, as the initial meson consists of one up-type and one down-type quark. 

\begin{figure}
    \centering
    \begin{subfigure}[b]{0.49\textwidth}
\begin{tikzpicture}
  \begin{feynman}
    \vertex (a) ;
    \vertex[right=1.5cm of a] (a2); 
    \vertex[right=1cm of a2] (b);
    \vertex at ($(b) + (1.1cm, -0.4cm)$) (a4); 
    \vertex at ($(a4) + (1.5cm, -0.55cm)$) (c) ;
    \vertex at ($(a) + (0cm, -1cm)$)(d) ;
    \vertex at ($(d) + (2.0cm, 0cm)$)(e) ;
    \vertex at ($(e) + (3.1 cm, -1.cm)$) (f);
    \vertex at ($(b) + (1.1 cm, 1.1cm)$) (g);
    \vertex at ($(g) + (1.5 cm, 0.5cm)$) (h);
    \vertex at ($(g) + (1.5 cm, -0.5cm)$) (i);
    
    \node[draw,ellipse,minimum width=1cm,minimum height=0.5cm, rotate=90,fill=gray!30,
  fill opacity=0.4] (Bc) at ($(a)!0.5!(d)$) {};
  \node[draw,ellipse,minimum width=1cm,minimum height=0.5cm, rotate=90,fill=gray!30,
  fill opacity=0.4] (B0) at ($(c)!0.5!(f)$) {};
  \node[draw,ellipse,minimum width=1cm,minimum height=0.5cm, rotate=90,fill=gray!30,
  fill opacity=0.4] (pip) at ($(h)!0.5!(i)$) {};
  \node[anchor=east, xshift=-0.25cm] at (Bc.center) {$B_c^+$};
  \node[anchor=west, xshift=0.25cm] at (B0.center) {$B^+$};
  \node[anchor=west, xshift=0.25cm] at (pip.center) {$\pi^0$};
  \vertex[right=0.9cm of Bc.center] (Wb) {$W$};  
  \vertex at ($(b) + (0cm, -0.4cm)$) (Dq) {$d,s,b$}; 
    
    \diagram* {
      {[edges=fermion]
        (a) -- [ edge label=\(c\)] (a2) -- (b)  -- (a4) -- [ edge label=\(u\)] (c), },
        (a4) -- [boson, out=-90, in=-90, looseness=1] (a2),
      (e) -- [fermion, edge label=\(\bar b\)] (d);
      (f) -- [fermion, edge label=\(\bar b\)] (e);
      (b) -- [photon, edge label=\({\gamma, Z}\)] (g);
      (g) -- [fermion, edge label=\(u\)] (h);
      (i) -- [fermion, edge label=\(\bar u\)] (g);
    };
  \end{feynman}
\end{tikzpicture}
    \caption{Colour-allowed EW penguin $P_{EW}$}
    \label{fig:placeholder}
    \end{subfigure}
    \begin{subfigure}[b]{0.49\textwidth}
\begin{tikzpicture}
  \begin{feynman}
    \vertex (a) ;
    \vertex at ($(a) + (1cm, 0 cm)$) (a1);
    \vertex at ($(a) + (2cm, 0 cm)$) (b);
    \vertex at ($(b) + (1.3cm, 0.5 cm)$) (b1);
    \vertex at ($(b) + (2.6cm, 1 cm)$) (c);
    \vertex at ($(a) + (0cm, -1cm)$)(d) ;
    \vertex at ($(d) + (2cm, 0cm)$)(e) ;
    \vertex at ($(e) + (2.6 cm, -1cm)$) (f);
    \vertex at ($(b) + (1.1 cm, -0.5cm)$) (g);
    \vertex at ($(g) + (1.5 cm, 0.5cm)$) (h);
    \vertex at ($(g) + (1.5 cm, -0.5cm)$) (i);
    \vertex at ($(b) + (0cm, 0.4cm)$) (Dq) {$d,s,b$};
    \node[draw,ellipse,minimum width=1cm,minimum height=0.5cm, rotate=90,fill=gray!30,
  fill opacity=0.4] (Bc) at ($(a)!0.5!(d)$) {};
  \node[draw,ellipse,minimum width=1cm,minimum height=0.5cm, rotate=90,fill=gray!30,
  fill opacity=0.4] (Bp) at ($(i)!0.5!(f)$) {};
  \node[draw,ellipse,minimum width=1cm,minimum height=0.5cm, rotate=90,fill=gray!30,
  fill opacity=0.4] (pi0) at ($(c)!0.5!(h)$) {};
  \node[anchor=east, xshift=-0.25cm] at (Bc.center) {$B_c^+$};
  \node[anchor=west, xshift=0.25cm] at (Bp.center) {$B_d^0$};
  \node[anchor=west, xshift=0.25cm] at (pi0.center) {$\pi^+$};
    
    \diagram* {
      (a) -- [fermion, edge label=\(c\)] (a1)-- [fermion] (b);
      (b) -- [fermion] (b1) -- [fermion, edge label=\( u\)] (c);
      (a1) -- [boson, out=90, in=120, looseness=1.1, edge label = \(W\)] (b1),
      (e) -- [fermion, edge label=\(\bar b\)] (d);
      (f) -- [fermion, edge label=\(\bar b\)] (e);
      (b) -- [gluon] (g);
      (h) -- [fermion, edge label'=\(\bar d\)] (g);
      (g) -- [fermion, edge label'=\(d\)] (i);
    };
  \end{feynman}
\end{tikzpicture}
\caption{QCD Penguin $P$}
\end{subfigure}
\caption{Subleading penguin diagrams.}
\label{fig: penguin diagrams}
\end{figure}

\begin{table}[t]
    \centering
    \renewcommand{\arraystretch}{1.25}
    \scalebox{0.85}{
        \begin{tabular}{lcccccc}
             \textbf{SCS Decay} & $P_{bD}$ & $P_{bD}^S$ & $P_{EW,bD}$ & $P_{EW,bD}^C$ & $A$\\
            \midrule
            \hline
             $B_c^+\to B^0\pi^+$ & $1$ & $0 $ & $0$ & $c_d $ & $1$\\[3mm]
             $B_c^+\to B^+\pi^0$ & $\frac{1}{\sqrt 2}$ & $0$ & $\frac{1}{\sqrt 2}(c_u-c_d)$ & $\frac{1}{\sqrt{2}}c_u$ & $\frac{1}{\sqrt 2}$\\[3mm]
             $B_c^+\to B^+\eta_q$ & $\frac{1}{\sqrt 2}$ & $\sqrt{2}$ & $\frac{1}{\sqrt{2}}(c_u+c_d)$ & $\frac{1}{\sqrt{2}}c_u$ & $\frac{1}{\sqrt 2}$ \\[3mm]
             \hline
             $B_c^+\to B_s^0K^+$ & $1$ & $0 $ & $0$ & $c_s $ & $1$\\[3mm]
             $B_c^+\to B^+\eta_s$ & $0$ & $1$ & $c_s $ & $0$ & $0$\\[3mm]
\hline \hline
        \end{tabular}
    }
    \caption{Topological amplitude decomposition for the CKM subleading amplitude of the SCS $B_c^+ \to B P$ decay modes proportional to $\lambda_b$. The upper part of the table correspond to $D=d$, while the lower one to $D=s$. $c_q$ represents the electric charge of the quark $q$.}
    \label{tab:BctoB_subleadtopologies}
\end{table}

\section{Mixing in Light Pseudoscalar and Vector Mesons}\label{ap:eta}

In Table~\ref{tab:Bc to B topologies}, we expressed the amplitudes in terms of the flavour eigenstates $\eta_q$ and $\eta_s$, which are defined as:
\begin{equation}
    \eta_q=\frac{1}{\sqrt{2}}(u\bar u + d\bar d) \hspace{1cm} \eta_s=s\bar s.
\end{equation}
The physical states $\eta$ and $\eta'$ arise from the mixing of these flavour eigenstates. We adopt the Feldmann–Kroll–Stech scheme, in which the $\eta_q-\eta_s$ system combines to form the physical states according to \cite{Feldmann:1998vh}:
\begin{equation}
\left(
\begin{array}{cc}
    \eta \\
    \eta'  \\
   \end{array}
\right)   
 = 
\left( 
   \begin{array}{cc}
    \cos\theta_P & -\sin\theta_P \\
    \sin\theta_P & \cos\theta_P \\
   \end{array} 
 \right)
 \left(
\begin{array}{cc}
    \eta_q \\
    \eta_s  \\
   \end{array}
\right).
\end{equation}
Consequently, the physical decay amplitudes are related to the flavour basis ones by:
\begin{equation}
    \begin{aligned}
        \mathcal{A}(B_c^+ \to B^+ \eta)&=\cos\theta_P \mathcal{A}(B_c^+ \to B^+ \eta_q) - \sin\theta_P\mathcal{A}(B_c^+ \to B^+ \eta_s) \\
        \mathcal{A}(B_c^+ \to B^+ \eta')&=\sin\theta_P\mathcal{A}(B_c^+ \to B^+ \eta_q)+\cos\theta_P\mathcal{A}(B_c^+ \to B^+ \eta_s)
    \end{aligned}
\end{equation}
The mixing angle $\theta_P$ is phenomenologically determined to be $\theta_P\simeq 41.6^\circ$ \cite{LHCb:2025sgp}.

The vector sector follows an analogous structure. The flavour eigenstates $\omega_q$ and $\omega_s$, which share the same flavour content as their pseudoscalar counterparts $\eta_q$ and $\eta_s$, mix to form the physical states $\omega$ and $\phi$. In this case, the mixing angle is very small, $\theta_V= 3.4^\circ$  \cite{Feldmann:2002kz}, which lies very close to the ideal limit in which the physical mesons are approximately,
\begin{equation}
    \omega\approx \frac{1}{\sqrt{2}}(u\bar u +d \bar d) \hspace{1cm} \phi \approx s\bar s.
\end{equation}

This near-ideal behaviour reflects the fact that, contrary to the pseucoscalar case, the vector meson mixing is not enhanced by the $U(1)_A$; instead, it originates primarily from OZI-suppressed gluonic transitions, which induce only a small deviation from ideal mixing \cite{Feldmann:2002kz}.

For the decays studied in this work, only the $\omega$ channel is kinematically accessible, since the available phase space is insufficient to produce a $\phi$ meson.

\section{Experimental \boldmath $\mathcal{B}(D\to M_1M_2)$ inputs in the data-driven approach}\label{ap:DMM_decays}
The experimental inputs used in the data-driven determination of the branching ratios are listed in Table~\ref{tab:exp_DtoMM}, together with the $B_c$ modes to which they are related.
\begin{table}[t]
\centering
    \renewcommand{\arraystretch}{1.5}
    \scalebox{0.9}{
    \begin{tabular}{ccc}
 $B_c^+ \to BM$ decay & Associated $D$ decay &  $\mathcal{B}(D\to M_1M_2)^{\mathrm{exp}}$ \\
\hline
\hline
$B_c^+ \to B^+ \bar K^0$ & $D_s^+ \to K^+ \bar K^0$ & $3.0 \times 10^{-2}$ \cite{Belle:2013isi}\\
$B_c^+ \to B^+ \bar K^{*,0}$ & $D_s^+\to K^+ \bar K^{*,0}$ & $0.13$\\
$B_c^+ \to B^0_s \pi^+$ & $D^0\to K^- \pi^+$ & $3.9 \times 10^{-2}$\\
$B_c^+ \to B^0_s \rho^+$ & $D^0\to K^- \rho^+$ & $0.11$\\
\hline
$B_c^+ \to B^+ \pi^0 $ & $D_s^+ \to K^+ \pi^0$ & $7.5 \times 10^{-4}$\\
$B_c^+ \to B^+ \eta $ & $D_s^+ \to K^+ \eta$ & $1.8 \times 10^{-3}$\\
$B_c^+ \to B^+ \eta' $ & $D_s^+ \to K^+ \eta'$ & $2.7 \times 10^{-3}$\\
$B_c^+ \to B^+ \rho^0 $ &$D_s^+ \to K^+ \rho^0$ & $2.2 \times 10^{-3}$\\
$B_c^+ \to B^+ \omega $ & $D_s^+ \to K^+ \omega $& $9.9 \times 10^{-4}$\\
$B_c^+ \to B_d^0 \pi^+ $ & $D^0 \to \pi^- \pi^+$ & $1.5 \times 10^{-3}$\\
$B_c^+ \to B_d^0 \rho^+ $ & $D^0 \to \pi^- \rho^+$ & $1.0 \times 10^{-2}$\\
$B_c^+ \to B_s^0 K^+ $ & $D^+ \to K_S^0 K^+$& $3.0 \times 10^{-3}$\\
\hline
\hline
\end{tabular}
}
\caption{Experimental inputs for $D \to M_1M_2$ branching ratios used in the determination of CF (upper) and SCS (under) $\mathcal{B}(B_c^+\to BM)$ decays. The values are taken from PDG \cite{ParticleDataGroup:2024cfk}, unless an explicit reference is provided. }
\label{tab:exp_DtoMM}
\end{table}

As discussed, for the modes marked with $\star$ in Table~\ref{tab:BctoB_BRs}, the partner $D$-meson channel is not measured. In these cases, the relevant branching ratios are adapted from other $B_c^+ \to BM$ decays, making use of $SU(3)_F$ symmetry or correcting for CKM factors where necessary. Specifically:
\begin{itemize}
    \item $\mathcal{B}(B_c^+\to B_{d(s)}^0 K^{*,+})$: is inferred from the pseudoscalar mode $B_c^+ \to B_{d(s)}^0 K^+$, employing an analogous factorization formula as in Eq.~\eqref{eq:BRDtoBc}.

    \item $\mathcal{B}(B_c^+\to B^+ K^{(*)0})$: is obtained from $\mathcal{B}(B_c^+\to B^+ \bar K^{(*)0})$ by correcting for CKM factors: $\mathcal{B}(B_c^+ \to B^+ K^{(*)0})=\left| \frac{V_{cd}^*V_{us}}{V_{cs}^*V_{ud}}\right|^2\mathcal{B}(B_c^+ \to B^+ \bar K^{(*)0})$

    \item $\mathcal{B}(B_c^+\to B_d^0 K^+)$: is related to $B_c^+ \to B_s^0 \pi^+$ by employing $SU(3)_F$ symmetry and applying the CKM factor correction $\left|\frac{V_{cd}^*V_{us}}{V_{cs}^*V_{ud}}\right|^2$.

\end{itemize}

As discussed in Sec.~\ref{sec:datadriven_approach}, the mapping used in the data-driven approach relies on the two decays sharing essentially the same topologies. Small differences arise from mismatches in annihilation contributions and from exchange and penguin-annihilation topologies present in $D$ decays, although these effects are expected to be negligible. Moreover, including such modes lies beyond our framework, as their factorization relation does not follow Eq.~\eqref{eq:fact}.

\bibliography{References2}

\end{document}